\documentclass[11pt]{article}

\usepackage{vmargin}
\setmarginsrb{3.2cm}{2.5cm}{3.2cm}{2.5cm}{1cm}{1cm}{1cm}{1.6cm}
\usepackage{amsmath}
\usepackage{amssymb}
\usepackage[dvips]{graphicx}
\usepackage{rotating}
\usepackage{psfrag}
\usepackage{fancyhdr}

\begin{document}
\begin{center}
{\LARGE \bf Chern-Simons Theory with Sources and\\  
Dynamical Quantum Groups I: Canonical Analysis and Algebraic Structures.
}

\bigskip
\bigskip
\bigskip

{ \sc E. Buffenoir\footnote{\tt Eric.Buffenoir@lpta.univ-montp2.fr},
 Ph.Roche\footnote{\tt Philippe.Roche@lpta.univ-montp2.fr Work supported by CNRS and EU under contract EUCLID
HPRN-ct-2002-00325.} 
 \\[1cm]}

{\it  UMR 5207 CNRS, Laboratoire de Physique Th\'eorique et d'Astroparticules, 
Universit\'e Montpellier II, Place Eug\`ene Bataillon,
 34095 Montpellier, France.
 }\\[3mm]

\vspace*{1.5cm}

\large{\bf Abstract}
\end{center}
\footnotesize
We study the quantization of Chern-Simons theory with group $G$ coupled to dynamical sources. 
We first study the dynamics of Chern-Simons sources in the Hamiltonian framework. The gauge group of this system is reduced to the Cartan subgroup of $G.$ We show that the Dirac bracket between the basic dynamical variables can be expressed in term of  dynamical $r-$matrix of rational type.
 We then couple minimally these sources to Chern-Simons theory with the use of a regularisation at the location of the sources. In this case, the gauge symmetries of this theory split in two classes, the bulk gauge transformation associated to the group $G$ and world lines gauge transformations associated to the Cartan subgroup of $G$. We give a complete hamiltonian analysis of this system and analyze in detail the Poisson algebras of functions invariant under the action of bulk  gauge transformations. This algebra is larger than the algebra of Dirac observables because it contains in particular functions which are not invariant under 
reparametrization of the world line of the sources.
We show that the elements of this  Poisson algebra have Poisson brackets expressed in term of  dynamical $r-$matrix of trigonometric type. This algebra is a   dynamical generalization of Fock-Rosly structure.
We analyze the quantization of these structures and describe different star structures on these algebras, with a special care to the case where $G=SL(2,{\mathbb R})$  and $G=SL(2,{\mathbb C })_{\mathbb R},$ having in mind to apply these results to the study of the quantization of 
massive spinning point particles  coupled to gravity with a cosmological constant in 2+1 dimensions.

\normalsize

\noindent

\section{Introduction}

 The aim of the present work is to provide new methods for the study of Chern-Simons theory associated to non-compact group coupled to dynamical sources.\\
  The combinatorial quantization program of Chern-Simons theory works perfectly well in the compact case as shown in \cite{AS} and for the case where $G=SL(2,{\mathbb C})_{\mathbb R}$ as shown in \cite{BNR}. This method provides a Hamiltonian quantization of Chern-Simons theory on $\Sigma_{g,n} \times {\mathbb R}$ where $\Sigma_{g,n}$ is a topological surface of genus $g$ with $n$ punctures: it provides a complete quantization of the algebra of Dirac Observables, i.e. the algebra of constants of motion of this theory as well as a unitary representation of it. \\
 Having in mind potential application to quantum gravity, this method needs further improvements to address "physical questions" related to the spectrum of important partial observables such as, for example, the radar time between two particles world-lines. The present work is a step in this direction.\\
In the second section of this paper, we study the dynamics of Chern-Simons sources in the Hamiltonian framework. The gauge group of this system is reduced to the Cartan subgroup of $G.$ We show that the Dirac bracket between the basic dynamical variables can be expressed in term of  dynamical $r-$matrix of rational type.
 In the third section, we then couple minimally these sources to Chern-Simons theory with the use of a regularisation at the location of the sources. In this case, the gauge symmetries of this theory split in two classes, the bulk gauge transformation associated to the group $G$ and world lines gauge transformations associated to the Cartan subgroup of $G$. We give a complete hamiltonian analysis of this system and analyze in detail the Poisson algebras of functions invariant under the action of bulk  gauge transformations. This algebra is larger than the algebra of Dirac observables because it contains in particular functions which are not invariant under
reparametrization of the world-line of the sources.
We show that the elements of this  Poisson algebra have Poisson brackets expressed in term of  dynamical $r-$matrix of trigonometric type. This algebra is a   dynamical generalization of Fock-Rosly structure.\\
In the fourth section, we analyze the quantization of the Poisson algebras involved in the study of free sources, and in the fifth section we analyze the quantization of the Poisson algebras involved in the study of sources coupled to Chern-Simons theory. In the study of these structures, we describe different star structures, with a special care to the case where $G=SL(2,{\mathbb R})$  and $G=SL(2,{\mathbb C })_{\mathbb R},$ having in mind to apply these results to the study of the quantization of
massive spinning point particles  coupled to gravity with a cosmological constant in $2+1$ dimensions.

\section{Hamiltonian Dynamic of a Chern-Simons Source.}
Let $G$ be a real connected simply-connected simple Lie group and ${\mathfrak g}$ be its real Lie algebra and denote $\langle,\rangle$ the Killing form on  $\mathfrak g$.

Let $\chi\in \mathfrak g$, 
we consider the action for a dynamical variable $g\in G$  defined as:

 \begin{equation}
S[g(t)]=\int _{t_{1}}^{t_{2}}dt\langle\chi,g^{-1}\frac{dg}{dt}\rangle\; \label{action}.\end{equation}

We will now proceed to the Hamiltonian analysis of this system. This system has been analyzed in the case where $G=SL(2,\mathbb{C})_{\mathbb R}$ in \cite{BN} where it describes the dynamic of a relativistic massive spinning particle on de Sitter space $dS_3.$

The cotangent bundle $T^* G$ is a trivial bundle and we have $T^* G\simeq\mathfrak g^*\times G.$ Let $X$ be an element of 
${ \mathfrak g}$, this defines a linear function on $\mathfrak g^*,$ which defines via the isomorphism of bundle a function 
on $T^* G$ denoted $f_{X}$.
If $\pi$ is a representation of $G$ acting on $V_{\pi}$, we  will view  its  matrix elements denoted $g^{\pi}$ as  functions on $T^* G$. 
The canonical symplectic structure on $T^* G$ induces the following Poisson brackets:
\begin{equation}
\{g^{\pi}_{1},g^{\pi'}_{2}\}=0\; \; ,\; \; \{f_{X} ,g^{\pi}\}={\pi}(X) g^{\pi}\; \; ,\; \; \{f_X,f_Y\}=f_{[X,Y]},\; \; \label{Poissonbracket1}\end{equation}
for all  $\pi,\pi'$  representations of $G$.

In the previous formulas, $g^{\pi}$ has to be viewed as an element of $End(V_{\pi}) \otimes Pol(G)$, where $Pol(G)$ is the Hopf algebra of matrix elements of finite dimensional representations of $G.$ We denote $Irr(G)$ the equivalence classes of irreducible finite dimensional representations of $G$.

$Pol(G)^*=\prod_{\pi \in Irr(G)} End(V_{\pi})$ is a (multiplier) Hopf algebra containing $U({\mathfrak g})$ as a Hopf subalgebra, therefore we can view $Pol(G)^*$ as a completion of $U({\mathfrak g})$ that we denote $U({\mathfrak g})^c.$ Any irreducible representation of $G$ defines a representation of $U({\mathfrak g})$ and then extends to a representation of $Pol(G)^*.$ 
 In order to obtain a  universal description, i.e independent of the choice of representation, we will collect all the $g^{\pi}$ in an object
$M\in U({\mathfrak g})^c\otimes Pol(G)$ such that
\begin{equation}(\pi \otimes id )(M)=g^{\pi}.\label{defM1}\end{equation}
By construction this  object satisfies:
\begin{equation}(\Delta \otimes id )(M)=M_1M_2,\label{defM2}\end{equation}
where $\Delta$ is the canonical coproduct on $Pol(G)^*=U({\mathfrak g})^c.$

The relations (\ref{Poissonbracket1}) can be equivalently written as:
\begin{equation}
\{M_{1},M_{2}\}=0\; \; ,\; \; \{P_{1},M_{2}\}=t_{12}M_{2}\; \; ,\; \; \{P_{1},P_{2}\}=[P_{1},t_{12}],\; \; \label{Poissonbracket}\end{equation}

where $t\in \mathfrak g^{\otimes 2}$ is defined by 
$t=\sum_{a,b}t^{ab} X_a\otimes X_b$ where $(X_a)$ is any basis of ${\mathfrak g}$,  $(t^{ab})$ is the inverse matrix of  $\langle X_a, X_b\rangle $  and $P:T^* G\rightarrow {\mathfrak g}, P=t^{ab}X_a f_{X_b}.$ As a remark, $P$ being in ${\mathfrak g}\otimes F(T^* G)$ , it obeys the relation $(\Delta\otimes id)(P)=P_1+P_2.$ 

This Hamiltonian system is a constrained system and the primary constraints 
are written as:
\begin{equation}
\phi=M^{-1}P M+\chi\approx 0.
\end{equation}
Note that $(\Delta \otimes id)(\phi)=\phi_1+\phi_2,$ as a result $\phi \in {\mathfrak g}\otimes F(T^* G).$  
If $\xi$ is an element of $ \mathfrak g$ we denote $\phi (\xi)=\langle \phi,\xi\rangle.$ 
The canonical hamiltonian is equal to zero, whereas the total Hamiltonian is equal to:
\begin{equation}
H^{tot}[M,P,\mu ]=\phi(\mu),\label{canonhamilt}\end{equation}
where $\mu$ is a Lagrange multiplier function with values in   $ \mathfrak g.$

The Poisson brackets of the constraints are
\begin{equation}
\{\phi_1,\phi_2\}=[\phi_2-\chi_2, t_{12}],
\end{equation} 
which can be recast as
\begin{equation}
\{\phi (\xi),\phi(\xi')\}=\phi([\xi,\xi'])-\langle\chi,[\xi,\xi']\rangle.
\end{equation}

Therefore we have:
\begin{equation}
\dot{\phi}=\{\phi,H^{tot}\}=[\mu,\phi-\chi].
\end{equation}

As a result,  conservation of the constraints under time evolution imposes no secondary
constraints and fixes $\mu $ to commute with $\chi.$ 
We denote $C_{\chi}$ the centralizer in $\mathfrak g$ of the element $\chi.$
For all $\xi\in C_{\chi},$ $\phi (\xi)$ is a first class constraint, and for every $\xi$ we have:
\begin{equation}
\delta_{\xi}M=\{M, \phi(\xi)\}=-M\xi\;\;,\;\;\delta_{\xi}P=\{P, \phi(\xi)\}=0.\label{bracketphiandm}
\end{equation}
The transformation of Lagrange multiplier, allowing for an invariance of the total action
 \begin{equation}
 S[M,P,\mu]=-\int _{t_{1}}^{t_{2}}dt \langle P,\frac{dM}{dt}M^{-1}\rangle+H^{tot}[M,P,\mu ]\; \label{action2},
 \end{equation}
 up to boundary terms, is given by
\begin{equation}
\delta_{\xi}\mu=\dot{\xi}+[\xi,\mu]. 
\end{equation}
Up to gauge symmetries which are trivial on-shell, the gauge symmetry corresponding to the reparametrization of the world line of the particle $t\mapsto t+\zeta(t)$ can be recovered from the gauge transformations generated by $\phi(\mu)\zeta(t).$ The Noether symmetry of this system is given by
\begin{equation}
M\mapsto g^{-1}M\;\;,\;\;P\mapsto g^{-1}Pg\;\;,\;\;\mu \mapsto \mu
\end{equation}
where $g$ is a constant element in $G.$

We will assume that $\chi$ is a regular semisimple element, therefore  ${\mathfrak h}=C_{\chi}$ is a real Cartan subalgebra.
We will now prefer to work with $\mathfrak{G}= \mathfrak{g}^{\mathbb C}.$
We have collected in the Appendix  1 the relevant conventions on complex  Lie algebras that we use throughout the rest of our work. 
We have also gathered, in Appendix 2, relevant notions about real structures. We will still denote by $M$ (resp. $P$) the complexification of $M$ (resp. $P$) and impose on it additional reality conditions imposed by the real form on $\mathfrak{G}$ selecting $\mathfrak{g}.$
Once a star $\star$ is defined on ${\mathfrak G},$ a star (denoted with the same symbol) can then be straightforwardly defined on $F(T^*G)$ by the following requirement:
\begin{equation}
({\star \otimes \star})(M)=M^{-1}, \;\;\;\;(\star\otimes \star) (P)=-P.\label{defM3}
\end{equation}
 As a result, coming back to the canonical analysis, the  action (\ref{action}) is real as soon as $\chi^{\star}=-\chi$. 

We now compute the Dirac bracket of the previous Hamiltonian system.
Assume that a constrained Hamiltonian system is given with a set of first class constraints $\varphi_i$ and  second class constraints $\psi_A.$ 
We will assume 
\begin{eqnarray}
\{\varphi_i,\varphi_j\}\approx 0\;\;\;\;\;\;\;\{\varphi_i,\psi_A\}\approx 0\;\;\;\;\;\;\;\{\psi_A,\psi_B\}={\cal D}_{AB}
\end{eqnarray}
where coefficients ${\cal D}_{AB}$ are functions on the phase space and ${\cal D}_{AB}$ is weakly invertible, i.e. there exist functions 
 $\Delta_{AB},(\Delta^{-1})^{AB},\nu_{AB}^C,\rho_{AB}^i$  on the phase space such that
\begin{eqnarray}
{\cal D}_{AB}=\Delta_{AB}+\nu_{AB}^C \psi_C+\rho_{AB}^i \varphi_i \;\;\;\;\;\;\; \Delta_{AB}(\Delta^{-1})^{BC}=\delta_A^C.
\end{eqnarray}
We define the Dirac bracket as follows:
\begin{equation}
\{f,g\}_{D}=\{f,g\}-\{f,\psi_A\}(\Delta^{-1})^{AB} \{\psi_B,g\}.
\end{equation}
The Dirac Bracket verifies the following properties, for any functions $f_1,f_2,f_3$ on the phase space:
\begin{eqnarray}
&&\{f_1,f_2\}_{D}=-\{f_2,f_1\}_{D}\;\;\;\;\;\;\; \{f_1,f_2f_3\}_{D}=\{f_1,f_2\}_{D}f_3+f_2\{f_1,f_3\}_{D},\\
&&\{f_1,\varphi_i\}_{D}\approx \{f_1,\varphi_i\}\; \; ,\; \;
\{f_1,\{\varphi_i,\varphi_j\}_{D}\}_{D}\approx \{f_1,\{\varphi_i,\varphi_j\}\}\;,\\
&&\{f_1,\psi_A\}_{D}\approx 0\; \; ,\label{fpsi0}\\
&&\{f_1,\{f_2,f_3\}_{D}\}_{D}+ cycl.perm. \approx 0\;\;{\rm if}\;\;\rho_{AB}^i\{\varphi_i,f_a\}=0\;\;,\forall a,A,B.
\end{eqnarray}
Due to the property (\ref{fpsi0}), one can strongly impose the constraints $\psi_A=0.$

Let ${\mathfrak G}={\mathfrak H}\oplus {\mathfrak H}^{\bot}$ where $ {\mathfrak H}^{\bot}=\mathfrak{N}^+\oplus\mathfrak{N}^-$ and for all $\xi\in {\mathfrak G}$ we denote by $\xi_{\mathfrak H}$
 (resp.$\xi_{{\mathfrak H}^{\bot}}$) its projection on ${\mathfrak H}$ 
(resp.${\mathfrak H}^{\bot}$). As a result we obtain:
\begin{equation}
\{\phi (\xi),\phi(\xi')\}\approx\langle \phi_{\mathfrak H }- \chi,[\xi,\xi']\rangle 
\approx -\langle ad(\tilde{\chi})(\xi),\xi'\rangle
\end{equation}
where we have   denoted $\tilde{\chi}=\chi-\phi_{\mathfrak H }.$

We denote  $Q( \tilde{\chi})$ the endomorphism of ${\mathfrak G}$ with kernel ${\mathfrak H}$ and which restriction on  ${\mathfrak H}^{\bot}$ is the inverse of  $-ad(\tilde{\chi})$ i.e  $Q( \tilde{\chi})(e_{\alpha})=
-\alpha(\tilde{\chi})^{-1}e_{\alpha}.$
We denote  $r( \tilde{\chi})\in {\mathfrak G}^{\wedge 2}$ by $ \langle \xi\otimes \xi',  r( \tilde{\chi})\rangle=- \langle  Q( \tilde{\chi})(\xi),\xi'\rangle$, it satisfies:
\begin{equation}
 r( \tilde{\chi})=-(\Delta^{-1})^{AB}\xi_{A}\otimes \xi_{B}\label{rdirac}
\end{equation}
where $(\xi_{A})$ is any  basis of  ${\mathfrak H}^{\bot}.$ 

The explicit expression of $r (\tilde{\chi})$ is given by: 
\begin{equation}
 r( \tilde{\chi})=-\sum_{\alpha\in\Phi }\frac{1}{\alpha(\tilde{\chi})}{e_{\alpha}\otimes e_{-\alpha}}.\label{rationalsolutionr}
\end{equation}
We denote $\tilde{\chi}=\sum_i \tilde{\chi}_{\alpha_i} \lambda^i$ with $\lambda^i$ the basis of 
$\mathfrak{H}$ dual to $(h_{\alpha_i})$ with respect to $\langle .,.\rangle,$ 
$r( \tilde{\chi})$ satisfies the classical dynamical Yang Baxter equation, i.e:
\begin{eqnarray}
&&[r_{12}(\tilde{\chi}), r_{13}(\tilde{\chi})+r_{23}(\tilde{\chi})]+
[r_{13}(\tilde{\chi}),r_{23}(\tilde{\chi})]=\nonumber\\
&&=
\sum_{i}(h_{\alpha_i}^{(1)}
\frac{\partial r_{23}(\tilde{\chi})}{\partial {\tilde{\chi}_{\alpha_i}}}
-h_{\alpha_i}^{(2)}\frac{\partial r_{13}(\tilde{\chi})}{\partial {\tilde{\chi}_{\alpha_i}}}+h_{\alpha_i}^{(3)}\frac{\partial r_{12}(\tilde{\chi})}{\partial {\tilde{\chi}_{\alpha_i}}}).
\end{eqnarray}
This solution is the basic rational dynamical r-matrix solution \cite{ES}.

As a result we get:
\begin{eqnarray}
&&\{M_{1},M_{2}\}_D=M_1 M_2 r_{12}( -\tilde{\chi}), \;\; \{M, \langle\tilde{\chi},h\rangle\}_D=M h,\; h\in {\mathfrak H}\nonumber\\
&& \{P_{1},M_{2}\}_D=t_{12}M_{2}\; \; ,\; \; \{P_{1},P_{2}\}_D=[P_{1},t_{12}].\; \; \label{DiracPoissonbracket}
\end{eqnarray}
This is straightforward from (\ref{Poissonbracket},\ref{bracketphiandm},\ref{rdirac}).

We can now impose strongly the second class constraints i.e $(M^{-1}PM+\chi)_{{\mathfrak H}^{\bot}}=0$, as a result we obtain the strong equality:
\begin{equation}
P=-M\tilde{\chi}M^{-1}.
\end{equation}

In  conclusion,
this Hamiltonian system is entirely described by $M, \tilde{\chi}=\sum_i \tilde{\chi}_{\alpha_i} \lambda^i$ with $\lambda^i$ the basis of 
$\mathfrak{H}$ dual to $(h_{\alpha_i})$ with respect to $\langle .,.\rangle,$ satisfying the relations:
\begin{eqnarray}
&&\{M_{1},M_{2}\}_D=M_1 M_2 r_{12}( -\tilde{\chi}), \;\;\;
\{M, \tilde{\chi}_{\alpha_i}\}_D=M h_{\alpha_i},\;\label{PoissonMdynamique1}\\
&&\{\tilde{\chi}_1,\tilde{\chi}_2\}_D=0,\;\;\;  \chi-\tilde{\chi}\approx 0,\label{PoissonMdynamique2}\\
&&H^{tot}=\langle \mu,\chi-\tilde{\chi}\rangle \;\;\;\;\;\;\;\;\;\;\;\;   (\Delta\otimes id )(M)=M_1M_2  .\label{PoissonMdynamique3}
\end{eqnarray}
We will denote by ${\cal S}(\mathfrak{G})$ the Poisson algebra
generated by $M, \tilde{\chi}$ and will call it the {\it Poisson algebra of the free source}.

We denote ${\cal S}(\mathfrak{G})^{\mathfrak H}$ the Poisson subalgebra of ${\cal S}(\mathfrak{G})$ Poisson commuting with the components of 
$\tilde{\chi}.$ This Poisson algebra is generated by the components of $P$.

We now specialize to the case where $\mathfrak{G}=sl(2,\mathbb{C})$ and 
denote $\tilde{\chi}=\tilde{\chi}_{0}\lambda.$
In the appendix 2, a brief summary of basic properties of star structures is given, in particular star involutions corresponding to the real forms of $sl(2,\mathbb{C})$ are given by equations (\ref{starsu2},\ref{starsu11},\ref{starsl2r}).
As said before, these star structures and automorphisms can be straightforwardly extended to functions on the phase space ${\cal S}(\mathfrak{G})$ by equation (\ref{defM3}). More precisely, if ${\buildrel {\frac{1}{2}} \over \pi}$ is the two dimensional unitary representation of $su(2)$, we define 
\begin{eqnarray}
&&g=({\buildrel {\frac{1}{2}} \over \pi} \otimes id)(M)=\left[ \begin{array}{cc}a & b\\ c & d\end{array}\right]
\end{eqnarray}
and the following linear automorphisms of ${\cal S}(\mathfrak{G}):$
\begin{eqnarray}
&&\sigma_1(a)=a\;\;\;\;\;\;\sigma_1(b)=-b\;\;\;\;\;\;\sigma_1(c)=-c\;\;\;\;\;\;\sigma_1(d)=d\;\;\;\;\;\;\sigma_1(\tilde{\chi}_0)=\tilde{\chi}_0\\
&&\sigma_2(a)=d\;\;\;\;\;\;\sigma_2(b)=-c\;\;\;\;\;\;\sigma_2(c)=-b\;\;\;\;\;\;\sigma_2(d)=a\;\;\;\;\;\;\sigma_2(\tilde{\chi}_0)=-\tilde{\chi}_0.
\end{eqnarray}
The corresponding star structures are then given by
\begin{eqnarray}
&&su(2):a^\star=d\;\;\;\;b^\star=-c\;\;\;\;\tilde{\chi}^\star=-\tilde{\chi},\;\;\;\;su(1,1): {\overline{\star}}=\sigma_1 \circ \star,\;\;\;\;sl(2,\mathbb{R}): {\underline{\star}}= \sigma_2 \circ \star.\nonumber\\&&\label{starclass1rk}
\end{eqnarray}
As a remark, note that  $\sigma_i \circ \star=\star \circ \sigma^{-1}_i$ ensuring that   these stars are  involutions.
We will denote ${\cal S}(\mathfrak{g})$ the algebra  ${\cal S}(\mathfrak{G})$ endowed with the corresponding star structure selecting the real form 
 $\mathfrak{g}.$

In order to apply our study to  the case of relativistic particles on a three dimensional deSitter space,  it is also
fruitful to analyze the action (\ref{action}) in the  slightly more general situation, i.e. $G=SL(2,\mathbb{C})_{\mathbb{R}}.$ We denote  $sl(2,\mathbb{C})_{\mathbb R}$  the real Lie algebra of $sl(2,\mathbb{C}),$  although this algebra is not simple, our whole construction can be straightforwardly generalized to this case. 
A complete analysis of the corresponding system, in this case, as well as the coupling to Chern-Simons theory has been done in \cite{BN}. 
The complexification  is such that $\mathfrak{g}^{\mathbb{C}}=sl(2,\mathbb{C})\oplus sl(2,\mathbb{C})$ and its Cartan basis will be denoted $e^{(l,r)},f^{(l,r)},h^{(l,r)}$ where
the $l$ (resp. $r$) generates the first (resp.second) component of the direct sum.
 Any of the star structures 
(\ref{starsl2c1},\ref{starsl2c2},\ref{starsl2c3}) can be used to recover the real Lie algebra
 $sl(2,\mathbb{C})_{\mathbb R}.$ 
Any real bilinear form on $sl(2,\mathbb{C})\oplus sl(2,\mathbb{C})$ invariant under the adjoint action is given by:
$<x,y>=a<x^{(l)},y^{(l)}>+\bar{a}<x^{(r)},y^{(r)}>$ with $a\in{\mathbb C}.$
Following \cite{BN}, we have chosen $a=\frac{1}{i}$ in order to describe the dynamic of a particle on deSitter space. This action will couple minimally to deSitter Gravity.
 The real action for the free source (\ref{action}) is 
entirely defined in terms of a Cartan element $\chi=\chi^{(l)}h^{(l)}+\chi^{(r)}h^{(r)}$ and depends on dynamical variables  $M^{(l)}, M^{(r)},\tilde{\chi}^{(l)},
\tilde{\chi}^{(r)}.$ The star structure on these elements can be easily described if we denote

 $$g^{(l)}=({\buildrel {(\frac{1}{2},0)} \over \pi} \otimes id)(M^{(l)})\;\;,\;\; g^{(r)}=({\buildrel {(0,\frac{1}{2})} \over \pi} \otimes id)(M^{(r)}),$$ indeed 
 \begin{eqnarray}
&&a^{(l)\bigstar}=d^{(r)}\;\;\;\;b^{(l)\bigstar}=-c^{(r)}\;\;\;\;d^{(l)\bigstar}=a^{(r)}\;\;\;\;c^{(l)\bigstar}=-b^{(r)}\;\;\;\;\tilde{\chi}^{(l)\bigstar}=-\tilde{\chi}^{(r)}\label{*sl2c1}\\
&&a^{(l)\overline{\bigstar}}=d^{(r)}\;\;\;\;b^{(l)\overline{\bigstar}}=c^{(r)}\;\;\;\;d^{(l)\overline{\bigstar}}=a^{(r)}\;\;\;\;c^{(l)\overline{\bigstar}}=b^{(r)}\;\;\;\;\tilde{\chi}^{(l)\overline{\bigstar}}=-\tilde{\chi}^{(r)}\label{*sl2c2}\\
&&a^{(l)\underline{\bigstar}}=a^{(r)}\;\;\;\;b^{(l)\underline{\bigstar}}=b^{(r)}\;\;\;\;d^{(l)\underline{\bigstar}}=d^{(r)}\;\;\;\;c^{(l)\underline{\bigstar}}=c^{(r)}\;\;\;\;\tilde{\chi}^{(l)\underline{\bigstar}}=\tilde{\chi}^{(r)}.\label{*sl2c3}
\end{eqnarray}

The corresponding  hamiltonian system is described by 
\begin{eqnarray}
&&\{M^{(\sigma)}_{1},M^{(\epsilon)}_{2}\}_D=M^{(\sigma)}_1 M^{(\sigma)}_2 r^{(\sigma\sigma)}_{12}( -\tilde{\chi}^{(\sigma)})\delta_{\sigma,\epsilon},\nonumber\\ 
&& \{M^{(\sigma)}, \tilde{\chi}^{(\epsilon)}\}_D=M^{(\sigma)} h^{(\sigma)}\delta_{\sigma,\epsilon},\;\;\;\;\;\;\;
\{\tilde{\chi}^{(\sigma)},\tilde{\chi}^{(\epsilon)}\}_D=0,\;\;\;\nonumber
\\
&&  \chi^{(\sigma)}-\tilde{\chi}^{(\sigma)}\approx 0\;\;\;\;\; H^{tot}=<\rho,(\chi-\tilde{\chi})>,\;\;\;\;\;\;\sigma,\epsilon \in\{l,r\}
\end{eqnarray}
where the Lagrange multipliers $\rho=\rho^{(l)}h^{(l)}+\rho^{(r)}h^{(r)}$ 
satisfy reality conditions $\rho^{(l)\star}=-\rho^{(r)}$ (resp.$\rho^{(l)\star}=\rho^{(r)} )$ in the case 
(\ref{*sl2c1},\ref{*sl2c2}) (resp. (\ref{*sl2c3})).

\section{ Chern-Simons theory coupled with  sources: Hamiltonian  analysis}
\label{sectionclassic}
\subsection{Hamiltonian reduction and Dirac Brackets}

We now proceed to the Hamiltonian analysis of  Chern-Simons theory coupled to classical sources. We have generalized the method developped in \cite{BN} to higher rank case.
 Let $\Sigma$ be an orientable topological compact surface of genus $g$ with $p$ punctures $x_1,...,x_p.$
We fix a Cartan subalgebra $\mathfrak{h}$ in $\mathfrak{g}.$
 We denote $\mathcal{M}=\Sigma\times [t_1,t_2],$ and to  each puncture $x_k$ we assign a regular semisimple  element $\chi_{(k)}\in {\mathfrak h}$.
The minimal coupling of Chern-Simons theory to pointlike sources located on the punctures would give the following action:
\begin{eqnarray}
S[A,M_{(1)},...,M_{(p)}]&&=\theta
\int_{\mathcal{M}}\langle A \wedge,  dA+\frac{2}{3}A\wedge A\rangle +\nonumber\\
&&+\sum_{k=1}^p\int_{t_1}^{t_2}dt \langle \chi_{(k)},M_{(k)}(t)^{-1}(\frac{d}{dt}+A_t(t,x_k)){M_{(k)}(t)}\rangle,
\end{eqnarray}
where $A$ is a ${\mathfrak g}-$connection on 
$\mathcal{M}$ and the real parameter $\theta$ is the coupling constant.\\

The naive field equations obtained from  the previous action are:
\begin{eqnarray}
\epsilon^{ij}(\partial_i A_j-\partial_j A_i+[A_i,A_j])=-\sum_{k=1}^p\frac{1}{4\pi \theta}{M_{(k)}\chi_{(k)}M_{(k)}^{-1}}\delta(x-x_k),
\end{eqnarray}
where $i,j$ run along spatial indices.

Because the right-handside contains $\delta$ distributions, $A$ has to belong to the set of distributions. In the non abelian case, the commutator term $A\wedge A$, involving the product of two distributions  is therefore ill-defined.
In order to solve this problem one can remove the points $x_k$ and impose, as in \cite{AM}, the condition $A_{\varphi}=-\frac{1}{4\pi \theta}{M_{(k)}\chi_{(k)}M_{(k)}^{-1}}$ as additional constraint on a small neighborhood of $x_k$ to the obvious bulk constraint $F=0,$ with the gauge group acting continuously at $x_k.$ The drawback of this method is that the constraint on $A_\varphi$ is imposed by hand and does not come from the canonical analysis of a well defined action principle.

In view of  these difficulties we will modify the previous action as explained in \cite{BN}.

\begin{figure}
\centering
\includegraphics[scale=0.7]{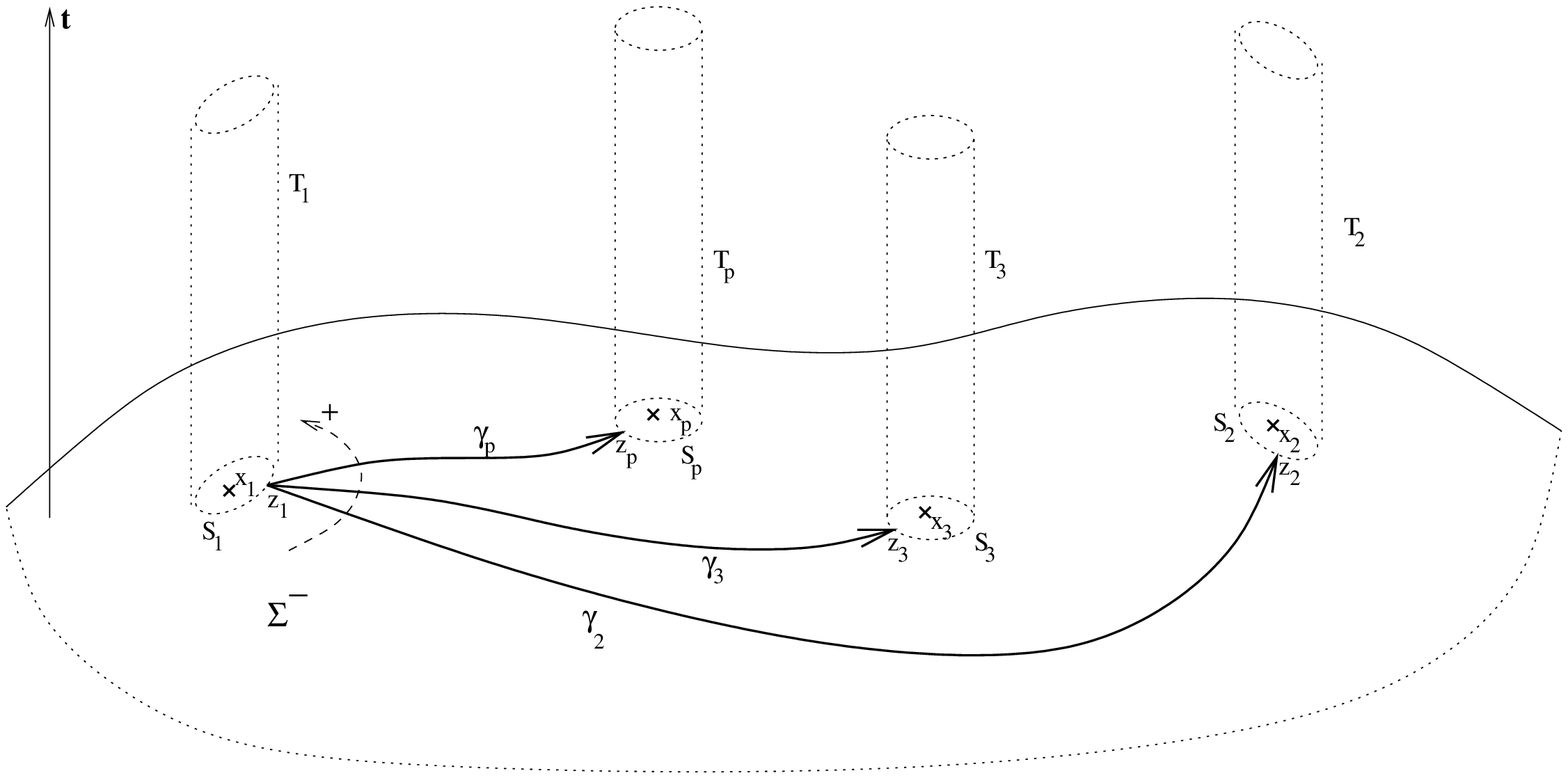}
\label{}
\end{figure}

To each puncture $x_k$ we associate a small closed  disk $D_k\subset \Sigma $ containing $x_k$ such that these disks do not intersect.
Let   $S_k$ be the closed curve defined by $S_k=\partial D_k$ and   denote by $T_k={\buildrel {\circ}\over {D_k}}\times [t_1,t_2]$ a tubular neighborhood of the vertical line passing through $x_k,$ and $T=\cup_k T_k.$
We denote  $\mathcal{M}^{-}=\mathcal{M}\setminus T,$ its boundary contains  ${B}_k= S_{k}\times [t_1,t_2]$ and 
$\Sigma^{-}_{t_1}\cup \Sigma^{-}_{t_2}$ where $\Sigma_t^{-}=\Sigma^-\times \{t \}$ with 
$\Sigma^{-}=\Sigma\setminus ( \cup_k{\buildrel {\circ}\over {D_k}})$.    The spatial boundary is ${\cal B}=\cup_{k=1}^{p} {B}_k.$ 
 $S_k$ being diffeomorphic to a circle, we choose a  parametrisation of $S_k$ by an angle $\varphi\in [0,2\pi[$ respecting the orientation and if $f$ is a function on $S_k$ we denote  $f^{av}$ its mean value, i.e:  $f^{av}=\frac{1}{2\pi}\int f(\varphi) d\varphi.$ For any function $f:\Sigma^- \rightarrow {\mathbb C}$ we denote $ f_{(k)}=f_{\vert S_k}.$
If $A$ is a connection on $\mathcal{M}^{-}$, $ S_{k}\times [t_1,t_2]$ being  embedded in  $\mathcal{M}^{-}$,  the pullback of  $A$ under this embedding gives a one form on ${B}_k$ denoted
  $A^{(k)}=A^{(k)}_t dt+ A^{(k)}_{\varphi}d\varphi.$ 
We will also denote  $S_t=\partial\Sigma_t^{-}=\cup_j S_t^j.$

We will study the following regularized action:
\begin{eqnarray}
&&S[A,M_{(1)},...,M_{(p)}]=\theta \int_{\mathcal{M}^{-}}\langle A \wedge,  dA+\frac{2}{3}A\wedge A\rangle+ \nonumber\\
&&+
\sum_{k=1}^p\int_{t_1}^{t_2}dt \langle \chi_{(k)},M_{(k)}^{-1}(\frac{d}{dt}+
(A^{(k)}_t)^{av}){M_{(k)}}\rangle+ \sum_{k=1}^p\theta \int_{{B}_k}\langle{A}^{(k)}_{t}|{A}^{(k)}_{\varphi}
 \rangle,
\end{eqnarray}
where $A$ is a smooth ${\mathfrak g}$-connection on ${\cal M}^-.$

We first analyze the gauge symmetry of this action.
We define 
\begin{equation}
 {\cal G}= \left\{g \in \mathcal{C}^{\infty }(\mathcal{M}^{-},G), 
\forall t\in [t_1,t_2], \;\;   g(.,t)\; \text{is a constant function on each}\; S_k\right\}.
\end{equation} 

The Lie algebra $\mathbb{g}$ of this group is:
\[
\mathbb{g}=\left\{ \Gamma \in \mathcal{C}^{\infty }(\mathcal{M}^{-},\mathfrak{g})/ \forall t\in [t_1,t_2], \;\;   \Gamma(.,t)\; \text{is a constant function on each}\; S_k\right\} \]

 and the action of $\xi \in \mathbb{g}$ on the gauge field $A$ and
on dynamical variables $M_{(k)}$, for $k\in \{1,\cdots ,p\}$ will
be respectively given by: \begin{eqnarray}
\delta[\Gamma ]A_{\mu }=D_{\mu }^{A}\Gamma \; \; \; , & \; \; \; \delta[\Gamma ]M_{(k)}=\Gamma_{\vert S_k}M_{(k)}\; \; \; ,\; \; \;  & \forall \Gamma \in \mathbb{g}.\label{symmetry}
\end{eqnarray}

As a result the infinitesimal action of the gauge group preserves the total action up to boundary terms on $\Sigma^{-}_{t_1}\cup \Sigma^{-}_{t_2}:$

\begin{equation}
\delta[\Gamma ]S[A,M_{(1)},...,M_{(p)}]=
\theta\int_{\Sigma _{t_{2}}^{-}}\! \! \! d^{2}x\, \epsilon^{ij}\langle A_{i},\partial_{j}\Gamma \rangle -\theta\int_{\Sigma _{t_{1}}^{-}}\! \! \! d^{2}x\, \epsilon ^{ij}\langle A_{i},\partial_{j}\Gamma\rangle. 
\end{equation}

We now study the Hamiltonian analysis of this action.
We reexpress the action as:

\begin{eqnarray}
S[A,M_{(1)},..,M_{(p)}] & = & \int \! dt\! \! \left(\theta\int \! \! \! \! \! \int _{\Sigma _{t}^{-}}d^{2}x\, \epsilon ^{ij}\langle A_{j}, \partial _{t}A{}_{i}\rangle+\langle A_{t},\epsilon ^{ij}F_{ij}(A)\rangle\right.\label{Lagrangian}\nonumber \\
 &  & \! \! \! \! \! \! \left.+\sum _{k=1}^{p}\langle\chi _{(k)},M_{(k)}^{-1}\frac{dM_{(k)}}{dt}\rangle+2\theta\sum_{k=1}^{p}\int _{S^k_{t}}\! \! \langle {A}_{t}^{(k)},{A}_{\varphi }^{(k)}-{X_{(k)}}\rangle\right)\label{actionwitten}
\end{eqnarray}
 where the mapping from the spatial
boundary $\mathcal{B}$ to $\mathfrak{g}$ denoted $X_{(k)}$ is defined by \begin{eqnarray}
X_{(k)} = -\frac{1}{4\pi \theta}{M_{(k)}\chi_{(k)}M_{(k)}^{-1}}.\label{definitionX1}
\end{eqnarray}

As in the free case we introduce the momenta $P_{(k)}\in \mathfrak{g}$ of the sources  which 
satisfy:
\begin{eqnarray}
\{M_{(k)\;1},M_{(l)\; 2}\}=0, & \{P_{(k)\; 1},M_{(l)\; 2}\}=t_{12}M_{(k)\;2}\delta _{kl}, & 
\{P_{(k)\;1},P_{(l)\;2}\}=[P_{(k)\;1},t_{12}]\delta _{kl},\nonumber \\
 &  & \label{PoissonbracketMP}
\end{eqnarray}
and the primary constraints are:
\begin{equation}
\phi_{(k)}=M_{(k)}^{-1}P_{(k)} M_{(k)}+\chi_{(k)}\approx 0,\;\;\;\;\; k=1,...,p.\label{phik}
\end{equation}

We denote by ${\cal C}_\phi$ this set of constraints.
$A_t$ is considered as a Lagrange multiplier and the Poisson structure on the spatial connection satisfies:
\begin{eqnarray}
\{A_{i}(x)_1,A_{j}(y)_2\}=\epsilon_{ij} t_{12}\; \delta ^{(2)}(x-y)\; , &  & \; \; \forall x,y\in \Sigma^{-}.\label{PoissonbracketAB}
\end{eqnarray}

Variation with respect to $A_t$ implements the primary constraint:

 \begin{eqnarray}
\Omega (v) & =  & \frac{1}{2}\int \! \! \! \! \! \int _{\Sigma^{-}}d^{2}x\; \epsilon ^{ij}\langle v,F_{ij}(A)\rangle+\sum_{k=1}^p\int _{S^k}\langle {v}_{(k)}, {A}_{\varphi }^{(k)}-{X_{(k)}}\rangle\approx 0\; ,\label{constraintOmega}\\
 &  & \; \forall v\in C^{\infty }(\Sigma^{-},\mathfrak{g}).\nonumber 
\end{eqnarray}

 We will denote by $\mathcal{C}_{\Omega }$ this set of constraints.

It has to be noticed that the previous constraints can be written as bulk constraints and boundary
constraints:\begin{eqnarray}
F_{ij}(x)=0 &  & \forall x\in \Sigma^{-}.\label{fixationinthebulk}\\
{A}_{\varphi }^{(k)}={X}_{(k)} &  & \label{fixationontheboundary}
\end{eqnarray}

 We deduce, from the action, the total Hamiltonian:
 \begin{eqnarray}
H^{tot}[A,M_{(k)},P_{(k)},\mu_{(k)} ,\rho ] & = & 
-\theta\left(\int \! \! \! \! \! \int _{\Sigma^{-}}d^{2}x\, \epsilon ^{ij}\langle \rho ,F_{ij}(A)\rangle+2\sum_{k=1}^p\int _{S_k}\! \! \langle \rho_{(k)},
 A_{\varphi }^{(k)}-X_{(k)}\rangle \right)\nonumber \\
 &  & +\sum _{k=1}^{p}\langle\mu_{(k)},\phi_{(k)}\rangle\; ,\label{hamiltoniantotalwitten}
\end{eqnarray}
where we have introduced  Lagrange multipliers $\mu_{(k)},\rho$ with $\mu_{(k)} \in \mathfrak{g}$,
and $\rho $ a smooth function on $\Sigma^-$  with value in $\mathfrak{g}$. 

Conservation
of the constraints (\ref{phik}) under time evolution imposes
the following conditions: \begin{eqnarray}
0\approx \frac{d\phi_{(k)}}{dt} & =
\left\{ \phi_{(k)},H^{tot}\right\}  & 
= [\mu_{{(k)}}-M_{(k)}^{-1}\rho_{{(k)}}M_{{(k)}},\chi_{(k)}]\; .\label{fixationofmu}
\end{eqnarray}

 As a result, the equations (\ref{fixationofmu}) do not impose any
secondary constraint.
We define $v_{(k)}=\mu_{(k)}-M_{{(k)}}^{-1}\rho_{{(k)}}M_{{(k)}}$, the previous equation imposes $v_{(k)}\in C_{\chi_k}.$

The requirement that the constraint $\Omega (v)$ must be preserved
in time implies no secondary constraints but imposes conditions on the Lagrange multiplier $\rho_{{(k)}}$ as now explained.  

Time evolution of $\Omega (v)$ is given by: \begin{eqnarray}
0\approx \frac{d\Omega (v)}{dt} & = & \Omega ([\rho,v])+\sum_{k=1}^p
\int _{S_k}\langle {v}_{(k)}, \partial _{\varphi }{\rho}_{(k)}+[X_{(k)},\rho_{(k)}-\rho_{(k)}^{av}]\rangle.\label{timeevolutionOmega}
\end{eqnarray}

For $\xi\in {\mathfrak g}$ we define the operator
\begin{equation}
K^{\xi}:C^{\infty }(S^1,\mathfrak{g})\longrightarrow C^{\infty }(S^1,\mathfrak{g}),\: {u}\mapsto \partial _{\varphi }{u}+[{\xi},{u}-u^{av}].
\end{equation}

Let $X\in \mathfrak{h},$ we will say that $X$ is special if $\exists \alpha\in \Phi ,e^{2\pi \alpha(X)}= 1.$

The condition on the Lagrange multiplier is:
\begin{equation}
K^{X_{(k)}}(\rho_{(k)})=0,
\end{equation} and 
this property has a very 
 simple meaning: in the case where $\frac{1}{4\pi\theta}\chi_{(k)}$ is not special, $\rho_{(k)}$ is constant on $S_k.$

Indeed from the equivariance property $K^{g\xi g^{-1}}\circ Ad_g =Ad_g\circ K^{\xi},$ it is sufficient to show that
 if   $X\in {\mathfrak h}$ is not special the kernel of $K^X$ is the set of constant functions on $S^1$.

To prove this, we define $G(\varphi)=
e^{\varphi X}(u-u^{av})e^{-\varphi X}.$
If $u$ lies in the kernel of  $K^X$,  $G(\varphi)$ is equal to a constant $G$. $2\pi$ 
periodicity of $u$ imposes that $G$ commutes with $e^{2\pi X}.$
The condition $e^{2\pi \alpha(X)}\not= 1$, ensures that $G_{{\mathfrak g}_{\alpha}}=0.$
Therefore $G\in {\mathfrak h}$ which implies, as announced, $u-u^{av}=G=0.$ 

We will assume in the sequel that every  $\frac{1}{4\pi\theta}\chi_{(k)}$ is not special.

The Dirac process ends and we are left with the set of constraints
$\mathcal{C}=\mathcal{C}_{\phi}\cup \mathcal{C}_{\Omega } $.

We replace the set of constraints $\mathcal{C}$ by $\mathcal{C}_{\phi}\cup \mathcal{C}_{\tilde{\Omega }}$
where 
\begin{eqnarray}
&&\tilde{X}_{(k)} = -\frac{1}{4\pi \theta}{M_{{(k)}}\tilde{\chi}_{{(k)}}M_{{(k)}}^{-1}}\label{definitionX2}\;\;\text{with}\;\; \tilde{\chi}_{{(k)}}=\chi_{(k)}-{\phi_{(k)}}_{\vert {\mathfrak h}},\\
&&\tilde{\Omega}(v)=\frac{1}{2}\int \! \! \! \! \! \int _{\Sigma^{-}}d^{2}x\; \epsilon ^{ij}\langle v,F_{ij}(A)\rangle +\sum_{k=1}^p\int _{S^k}\langle {v}_{(k)}, {A}_{\varphi }^{(k)}-{\tilde{X}_{(k)}}\rangle .
\end{eqnarray}

In order to compute the Dirac bracket we proceed iteratively.

 We treat first the subset of second class constraints in $\mathcal{C}_{\phi}$ as in the first section and compute the resulting  intermediary Poisson 
bracket denoted $\{\cdot,\cdot\}_d:$
\begin{eqnarray}
&&\{M_{{(k)}\;1},M_{{(l)}\;2}\}_d=M_{{(k)}\; 1} M_{{(k)}\; 2} r_{12}( -\tilde{\chi}_{(k)})\delta_{kl},
 \;\;\;\{M_{(k)}, \tilde{\chi}_{(l)\alpha_i}\}_d=M_{(k)} h_{\alpha_i} \delta_{kl},\nonumber\\
&&\{\tilde{\chi}_{{(k)}\;1},\tilde{\chi}_{{(l)}\;2}\}_d=0,\;\;\;  \chi_{(k)}-\tilde{\chi}_{(k)}\approx 0,\;\;P_{(k)}=-M_{(k)}^{-1} \tilde{\chi}_{(k)} M_{(k)}.
\end{eqnarray}
The Poisson brackets involving the connection are left unchanged.

We now compute the Poisson brackets of the remaining constraints:
 \begin{eqnarray}
&&\{\tilde{\Omega }(u),\tilde{\Omega }(v)\}_{d}=
\frac{1 }{2\theta}\left(\tilde{\Omega }([u,v])-\sum_{k=1}^p
\int _{S_k}\langle {u},K^{\tilde{X}_{(k)}}{v}\rangle \right) \label{PoissonbracketOmegaOmega}\\
&&\{\tilde{\chi}_{(k)}-\chi_{(k)},\tilde{\Omega }(u)\}_d=0 \;\;\;  \;\forall u,v\in C^{\infty }(\Sigma ^{-},\mathfrak{g}).
\end{eqnarray}

 We now have to distinguish first class from second class constraints.

{}From the Poisson brackets (\ref{PoissonbracketOmegaOmega}), we see that,
given $u\in C^{\infty }(\Sigma ^{-},\mathfrak{g})$, the constraint
$\tilde{\Omega }(u)$ is first class if and only if $u_{\vert S_k}$ is constant 
for each $k.$

Moreover the constraints $\tilde{\chi}_{(k)}-\chi_{(k)}$ are also first class.

There is no canonical way to select the set of second class
constraints, however, the Dirac bracket does not depend on this choice.
This procedure is achieved by choosing  a space of functions ${\mathbb{g}^{\bot}}$
such that $C^{\infty }(\Sigma^{-},\mathfrak{g})=\mathbb{g}\oplus {\mathbb{g}^{\bot}},$
and $\forall u\in {\mathbb{g}^{\bot}},\: u^{av}_{\vert S_k}=0$ for any $k=1,...,p.$ The detail
of this space in the bulk is in fact irrelevant. The key property of this space is the fact that for any couple $u,v \in {\mathbb{g}^{\bot}},$ if $\forall k=1,...,p, u_{\vert S_k}=v_{\vert S_k}$ then $u=v.$ As a result we can identify ${\mathbb{g}^{\bot}}$ with the vector space of functions of zero mean value from the boundary circles $\cup_{k=1}^p S_k$ to the Lie algebra. Hence, it can be naturally equipped with a pre-Hilbert space structure using the $L^2$ norm.\

In order to compute the final Dirac bracket, it will be useful to introduce
the antisymmetric bilinear form ${\cal K}^{\tilde{X}}$ as follows: 
\begin{eqnarray}
{\mathbb{g}^{\bot}}\, ^{\times 2} & \longrightarrow  & \mathbb{C}\nonumber \\
(u,v) & \longmapsto  &{\cal K}^{\tilde{X}} (u,v)=\sum_{k=1}^p\int _{S_k}\langle {u}_{(k)},K^{\tilde{X}_{(k)}}{v}_{(k)}\rangle.
\end{eqnarray}
 Given an element $u\in {\mathbb{g}^{\bot}}$, ${\cal K}^{\tilde{X}}(u,\cdot ):{\mathbb{g}^{\bot}}\longrightarrow \mathbb{C}$
is invertible and we  denote  $({\cal K}^{\tilde{X}}){}^{-1}(\cdot ,u)$ its inverse,
i.e. \begin{eqnarray}
 \int _{ {\mathbb{g}^{\bot}}}[\mathcal{D}w]{\cal K}^{\tilde{X}}(u,w)({\cal K}^{\tilde{X}})^{-1}(w,v)=\delta_{{\mathbb{g}^{\bot}}}(u-v)\; , &  & \forall \; u,v\in {\mathbb{g}^{\bot}}\; 
\end{eqnarray}
where the previous functional integration is defined via the  integration  on the Fourier modes of functions on  $\cup_{k=1}^p S_k.$

{}From the expression (\ref{PoissonbracketOmegaOmega}), the Poisson bracket of two second class constraints  is strongly   equal   to the sum of three terms 
\begin{equation}
 2\theta\{\tilde{\Omega }(u),\tilde{\Omega }(v)\}_{d}=\tilde{\Omega }([u,v]_{\vert {\mathbb{g}^{\bot} }})+
\tilde{\Omega }([u,v]_{\vert \mathbb{g} })-{\cal K}^{\tilde{X}}(u,v).
\end{equation}

To obtain the expression of  the Dirac matrix it is sufficient to compute it up to second class constraints: we can therefore eliminate the first term appearing in the previous expression.
Moreover, to compute the expression of the Dirac bracket on functions which are invariant under the first class constraints $\tilde{\Omega }(w)$ for
$w\in \mathbb{g},$ it is sufficient to invert the Dirac matrix computed only up to these first class constraints.
As a result we obtain that:

\begin{eqnarray}
\{f,g\}_{D} & = & \{f,g\}_{d}+{2\theta}\int \! \! \! \! \! \int _{{\mathbb{g}^{\bot}}\times {\mathbb{g}^{\bot}}}[\mathcal{D}u][\mathcal{D}v]\{f,\tilde{\Omega }(u)\}_{d}({\cal K}{}^{\tilde{X}}){}^{-1}(u,v)\{\tilde{\Omega }(v),g\}_{d}\label{definitiveDiracbracket}
\end{eqnarray}
for any $f,g$ which Poisson commute with  
 $\tilde{\Omega }(w), w\in\mathbb{g}. $

Let us now  consider three different punctures $x_l, x_m, x_n$ and let  $\gamma$ (resp. $\gamma'$) be an oriented  curve joining $S_l$ to $S_m$ (resp. $S_l$ to $S_n$). We denote by $\varphi$ and   $\varphi'$ the angles associated respectively to the departure point of  $\gamma$ and  $\gamma'$. 

Let $U_\gamma$ be the holonomy of the connection $A$ along $\gamma$.
The functions $V_\gamma[A,M]=M_{(m)}^{-1}U_{\gamma}M_{(l)}$ are invariant under gauge transformation belonging to ${\cal G}$.
We can compute explicitely the Dirac bracket of these functions, computation done in the Appendix 4, which can be expressed 
as a dynamical quadratic Poisson bracket as
\begin{eqnarray}
\{V_{\gamma 1},V_{\gamma' 2}\}_{D}=V_{\gamma 1}V_{\gamma' 2} r^{\theta}_{12}
(\varphi-\varphi';-\tilde{\chi}_{(k)})\label{poissonbracketofVr}
\end{eqnarray}
with
$ r^{\theta}_{12}(\varphi;\tilde{\chi})$ given by:

\begin{equation}
r^{\theta}_{12}(\varphi;-\tilde{\chi})=\frac{1}{4\pi \theta}\left( (\pi-\varphi)\sum_{j}h_{\alpha_j}\otimes \lambda^j+\sum_{\alpha\in \Phi}
e_\alpha\otimes e_{-\alpha}\frac{\pi e^{{\tilde{\chi}}(\alpha)(\pi-\varphi)/4\pi\theta}}{\sinh ({\tilde{\chi}}(\alpha)/4\theta)}\right).
\label{expressionofrtrigodynamique}\end{equation}

Note that there is a discontinuity of this function at $\varphi=0$ and that 
$ r^{\theta}_{12}(0+;-\tilde{\chi})=r^{\theta\; (+)}_{12}(-\tilde{\chi})$ and
$ r^{\theta}_{12}(0-;-\tilde{\chi})=r^{\theta\; (-)}_{12}(-\tilde{\chi})$ are the
basic trigonometric  solutions of the classical dynamical Yang-Baxter 
equation which satisfy
\begin{equation}
 r^{\theta\; (+)}_{12}(-\tilde{\chi})-r^{\theta\; (-)}_{12}(-\tilde{\chi})=
\frac{1}{2\theta}t_{12}.
\end{equation}
We will denote 
$r^{\theta\; (\pm)}_{12}=r^{\theta\; (\pm)}_{12}(\tilde{\chi}\rightarrow-\infty)$ and note that $\theta r^{\theta\; (\pm)}_{12}$ is the standard 
solution of the classical Yang-Baxter equation.

\subsection{Poisson algebras of dynamical  holonomies}

Let us stress that the function $V_\gamma$ is  not a Dirac  observable 
because it does not  Poisson commute with $\tilde{\chi}_m,\tilde{\chi}_l$ and we have:

\begin{eqnarray}
\{V_\gamma, \tilde{\chi}_{(l)\alpha_i} \}=V_\gamma h_{\alpha_i}\;\;,\;\;
\{V_\gamma, \tilde{\chi}_{(m)\alpha_i} \}=-h_{\alpha_i} V_\gamma.\;\;\label{PoissonVeth}
\end{eqnarray}

Let us fix for each $S_k$ a point $z_k$ on it, and fix $p-1$ curves
 $\gamma_{(2)},...,\gamma_{(p)}$ where $\gamma_{(k)}$ goes from $z_1$ 
to $z_k$. We will assume that $\gamma_{(k)}$ does not touch any circle 
except its end points and that two different curves have only $z_1$ as 
intersection. We will study
the Poisson algebra of 
polynomials of matrix elements of the holonomies
$(V_{\gamma_{(k)}})_{k=2,\cdots, p}$ with coefficients in the algebra
of functions of $(\tilde{\chi}_{(k)})_{k=1,\cdots, p}.$
The orientation of the surface $\Sigma$ at the point $z_1$ fixes an order 
$<$ on the set of curves  $\gamma_{(2)},...,\gamma_{(p)}.$
Up to relabelling of the sources it is always possible to assume that
  $\gamma_{(2)}<\cdots<\gamma_{(p)}.$

The Poisson
brackets between these elements are then given by:
\begin{eqnarray}
&&\{V_{\gamma_{(k)}}, \tilde{\chi}_{(1)\alpha_i} \}=V_{\gamma_{(k)}} h_{\alpha_i}\;\;,\;\;
\{V_{\gamma_{(k)}}, \tilde{\chi}_{(l)\alpha_i} \}=-h_{\alpha_i}
 V_{\gamma_{(k)}}\delta_{kl},\;\;\\
&&\{V_{\gamma_{(k)} 1},V_{\gamma_{(l)} 2}\}=
V_{\gamma_{(k)} 1}V_{\gamma_{(l)} 2} r^{\theta (-)}_{12}
(-\tilde{\chi}_{(1)})\;\;\text{with}\;\;k<l,
\label{poissonV1}\\
&&\{V_{\gamma_{(k)} 1},V_{\gamma_{(k)} 2}\}=V_{\gamma_{(k)} 1}
V_{\gamma_{(k)} 2} r^{\theta (-\epsilon)}_{12}(-\tilde{\chi}_{(1)}) + 
r^{\theta (\epsilon)}_{12}(-\tilde{\chi}_{(k)})V_{\gamma_{(k)} 1}
V_{\gamma_{(k)} 2},\label{poissonV2}\\
&&(\Delta\otimes id)(V_{\gamma_{(k)}})=V_{\gamma_{(k)\;1}}V_{\gamma_{(k)\;2}}
\end{eqnarray}

We denote $Hol_p(\mathfrak{G})$ this Poisson algebra and call it the {\it Poisson algebra of dynamical boundary-boundary holonomies}.

Remark: We could have generalized straightforwardly  the function $V_\gamma[A,M]$ to embedded 
spin networks with open legs attached to the circles $S_k.$
The Poisson bracket of these objects have  two types of contributions:
those associated to intersecting points in the bulk which are computed using Goldman bracket and 
those associated to the boundary which are computed using (\ref{poissonV1},\ref{poissonV2}).

In the  case where $\Sigma$ is the sphere a similar construction based on spin-networks is equivalent to the previous one.
In the case where $\Sigma$ is a genus $n$-surface the complete description of observables requires the
introduction of holonomies corresponding to non-trivial cycles. This description can be developped
straightforwardly along the guidelines of previous derivations. It is not the aim of the present
paper to develop these aspects, we will then stick to the case where  $\Sigma$ is the sphere. 

At this point, we still have to impose the remaining relations
corresponding to the requirement of flatness of the connection on the sphere. This requirement is implemented by the relation
\begin{eqnarray}
&& \Upsilon-1\approx 0, \label{flat1}
\end{eqnarray}
with
\begin{eqnarray}
&& \Upsilon=e^{-\frac{\tilde{\chi}_{(1)}}{2\theta}}\prod^2_{k=p}\left( V_{\gamma_{(k)}}^{-1}
e^{-\frac{\tilde{\chi}_{(k)}}{2\theta}}V_{\gamma_{(k)}}\right).
\end{eqnarray}
It is important to emphasize that these relations are not remnants of some large bulk gauge transformations $\tilde{\Omega}$ ($\Upsilon$ is even strictly invariant under $\tilde{\Omega}$ !). These relations have to be considered as additional relations to be implemented on $V_{\gamma_{(k)}},\tilde{\chi}_{(k)}$ in order to recover the true degrees of freedom and not as canonical constraints generating any gauge transformations. 
We have to note that, using previous Poisson brackets as well as the relation
\begin{eqnarray}
&&r^{\theta\;(+)}_{12}(-\tilde{\chi}_{(k)})-\frac{1}{2\theta}\sum_j h_{\alpha_j}\otimes\lambda^j=
e^{\frac{\tilde{\chi}_{(k)}}{2\theta}} r^{\theta\;(-)}_{12}(-\tilde{\chi}_{(k)}) e^{-\frac{\tilde{\chi}_{(k)}}{2\theta}},
\end{eqnarray}
the Poisson brackets between dynamical variables and $\Upsilon$ can be straighforwardly computed and are:
\begin{eqnarray}
&& \{\Upsilon_1,V_{\gamma_{(k)}\;2}\}=V_{\gamma_{(k)}\;2}[\Upsilon_1,r^{\theta (-)}_{12}(-\tilde{\chi}_{(1)})]\\
&&\{\Upsilon,\tilde{\chi}_{(1)\;\alpha_i}\}=[\Upsilon,h_{\alpha_i}],\;\;\;\;\;\;\;\;\;\;\{\Upsilon_1,\tilde{\chi}_{(k)}\}=0,\;\;
\forall k=2,...,p.
\end{eqnarray}
As a result $(\Upsilon-1)Hol_p(\mathfrak{G})$ is a Poisson ideal, therefore we can define 
  the Poisson algebra $Hol_p(S^2,\mathfrak{G})$ obtained from $Hol_p(\mathfrak{G})$ by moding out the relations $\Upsilon=1.$ 
  We call $Hol_p(S^2,\mathfrak{G})$ the {\it algebra of dynamical boundary-boundary holonomies on $S^2$ of zero curvature.}

In order to implement these constraints in a way adapted to quantization, we will give an alternative description of 
$Hol_p(S^2,\mathfrak{G})$ by embedding  $Hol_p(\mathfrak{G})$ in a
larger Poisson 
algebra $Hol_p(\bullet,\mathfrak{G})$,   the {\it algebra of bulk-boundary dynamical holonomies } defined as follows:
 $Hol_p(\bullet,\mathfrak{G})$ is generated by
polynomials of matrix elements of the matrices   
$(U_{(k)})_{k=1,\cdots, p}$ with coefficients in the algebra
of functions of $(\tilde{\chi}_{(k)})_{k=1,\cdots, p},$ with Poisson brackets
\begin{eqnarray}
&&\{U_{(k)}, \tilde{\chi}_{(l)\alpha_i} \}=
U_{(k)} h_{\alpha_i}\;\;,\label{poissongravalex1}\\
&&\{U_{(k) 1},U_{(l) 2}\}=
\breve{r}^{\theta (-)}_{12} U_{(k) 1}U_{(l) 2} \;\;\text{with}\;\;k<l
\label{poissongravalex2}\\
&&\{U_{(k) 1},U_{(k) 2}\}=
U_{(k) 1}U_{(k) 2}
r^{\theta (-\epsilon)}_{12}(-\tilde{\chi}_{(1)}) +
\breve{r}^{\theta (\epsilon)}_{12}
U_{(k) 1}U_{(k) 2},\label{poissongravalex3}\\
&&(\Delta\otimes id)(U_{(k)})=U_{(k)1}U_{(k)2}.
\end{eqnarray}
In order for this Poisson algebra to be well defined the $r-$matrix $\breve{r}^{\theta}$ has to 
be a solution of classical Yang-Baxter equation such that 

\begin{equation}
\breve{r}^{\theta\; (+)}_{12}-\breve{r}^{\theta\; (-)}_{12}=
\frac{1}{2\theta}t_{12}.\label{conditiononrtilde}
\end{equation}

The Poisson-Lie group $F(G^{\mathbb C})$ which Poisson bracket is defined by
\begin{eqnarray}
&&\{g_1,g_2\}=[g_1g_2,\breve{r}^{\theta}_{12}]
\end{eqnarray}
coacts on $Hol_p(\bullet,\mathfrak{G})$ by the Poisson map
\begin{eqnarray}
Hol_p(\bullet,\mathfrak{G})  &\longrightarrow& F(G^{\mathbb C}) \otimes Hol_p(\bullet,\mathfrak{G}) \\
U_{(k)} &\longrightarrow&
g^{-1}U_{(k)},
\end{eqnarray}
where the Poisson structure on $F(G^{\mathbb C}) \otimes Hol_p(\bullet,\mathfrak{G}) $ is defined as
$\{f\otimes a, f'\otimes a'\}=\{f, f'\}\otimes a a'+f f'\otimes \{ a,  a'\}.$

As a result the  coinvariant elements of  $Hol_p(\bullet,\mathfrak{G})$ is a Poisson subalgebra 
$Hol_p(\bullet,\mathfrak{G})^{\mathfrak{G}}$ which is moreover the image of the following injective Poisson map:
\begin{eqnarray}
Hol_p(\mathfrak{G}) &\longrightarrow& Hol_p(\bullet,\mathfrak{G}) \nonumber\\
(V_{\gamma_{(k)}},\tilde{\chi}_{(k)},\tilde{\chi}_{(1)}) &\mapsto&
(U^{-1}_{(k)}U_{(1)},\tilde{\chi}_{(k)}, \tilde{\chi}_{(1)}), \;\;\; k\geq 2.
\end{eqnarray}

It can easily be shown that different choices of $r-$matrix $\breve{r}^{\theta}$ fullfilling the required condition 
(\ref{conditiononrtilde}) give rise to the same Poisson algebra of gauge coinvariant elements. We will then generically choose
  $\breve{r}^{\theta}=r^{\theta}$ (however, in the $sl(2,{\mathbb C})_{\mathbb R}$ case, a different choice 
  will be more fruitful for our purpose). \\

Our aim is now to implement the flatness condition (\ref{flat1}) in $ Hol_p(\bullet,\mathfrak{G}).$ We define
\begin{eqnarray}
&&\Gamma=\prod^1_{k=p}\left( U_{(k)}
e^{-\frac{\tilde{\chi}_{(k)}}{2\theta}}U_{(k)}^{-1}\right).\label{flat2}
\end{eqnarray}
It can easily be shown that the elements
Poisson commuting with $\Gamma$ are exactly the gauge coinvariant elements in $Hol_p(\bullet,\mathfrak{G}).$ As a result we 
obtain $Hol_p(\bullet,\mathfrak{G})^\mathfrak{G}=Hol_p(\mathfrak{G}).$

The elements of $Hol_p(\bullet,\mathfrak{G})$ Poisson commuting with $\Gamma$ modded by the relations $\Gamma=1$ is endowed 
with a natural Poisson algebra which is  isomorphic to the  Poisson algebra $Hol_p(S^2,\mathfrak{G}).$  

It will be  useful, in order to deal with the invariance under the remaining first class constraints:
\begin{eqnarray}
&&\pi_{(k)}=\tilde{\chi}_{(k)}-\chi_{(k)}\approx 0,\;\;k=1,...,p\label{constraintchi}
\end{eqnarray}
to introduce the following objects 
\begin{eqnarray}
&&{\cal P}_{(k)}=U_{(k)}e^{-\frac{\tilde{\chi}_{(k)}}{2\theta}}U_{(k)}^{-1}
\end{eqnarray}
generating the Poisson subalgebra of elements invariant under the remaining first class constraints $\pi_{(k)}.$
The Poisson algebra generated by $({\cal P}_{(k)})$ is the  classical multi-loop Poisson 
algebra ${\cal L}_{0,p}(\mathfrak{G})$ of
 Fock-Rosly \cite{FR}.
If $\pi$ is a finite dimensional representation of $\mathfrak{G},$ we define $C^{\pi}_{(k)}=(tr_{\pi}\otimes id)({\cal P}_{(k)}),$ 
which are Poisson central elements of ${\cal L}_{0,p}(\mathfrak{G}).$
For $\chi_{(1)},...,\chi_{(p)}$ we denote
$I_{\chi_{(1)},...,\chi_{(p)}}$ the ideal generated by $C^{\pi}_{(k)}-tr_{\pi}(e^{-\frac{{\chi}_{(k)}}{2\theta}}),$ 
 $\forall\pi, \forall k.$

We then define the Poisson algebra ${\cal L}_{0,p}(\mathfrak{G};\chi_{(1)},...,\chi_{(p)})=
{\cal L}_{0,p}(\mathfrak{G})/I_{\chi_{(1)},...,\chi_{(p)}}.$

We can now define the Poisson algebra of elements of $Hol_p(S^2,\mathfrak{G})$ Poisson commuting
 with all the remaining 
first class constraints  ${(\pi_{(k)})}_{k=1,...,p}$ modded out by the relation $\pi_{(k)}\approx 0,\;\;k=1,...,p.$
This Poisson algebra is isomorphic to the moduli space of flat connection on $S^2$ with p-punctures associated to 
conjugacy classes which representative are  $e^{-\frac{{\chi}_{(k)}}{2\theta}}.$
This Poisson algebra is usually denoted $M(S^2,\mathfrak{G};\chi_{(1)},...,\chi_{(p)})$ and is 
obtained from the 
Poisson algebra ${\cal L}_{0,p}(\mathfrak{G})$ by modding out the flatness condition, i.e.
$$M(S^2,\mathfrak{G};\chi_{(1)},...,\chi_{(p)})={\cal L}_{0,p}(\mathfrak{G};\chi_{(1)},...,\chi_{(p)})^\mathfrak{G}/(\Gamma=1).$$

Note that we have worked with the complexification of $\mathfrak{g}$ througout this section. One can define the analogous Poisson
algebras $Hol_p(\mathfrak{g}), Hol_p(\bullet,\mathfrak{g}),Hol_p(S^2,\mathfrak{g}),$  by defining on 
$Hol_p(\mathfrak{G}),$ $ Hol_p(\bullet,\mathfrak{G}),$ $ Hol_p(S^2,\mathfrak{G}),$ star structures selecting the real form
$\mathfrak{g}.$ We are sketchy at this point but we will develop this in the case of the quantization of these Poisson algebras.

The  Dirac quantization program has been successfully constructed in \cite{AS}
for the case of the Poisson algebra  $M(S^2,\mathfrak{G};\chi_{(1)},...,\chi_{(p)})$ through the use of the multiloop algebra and 
its representations. In this framework, all the first class gauge constraints are promoted to projectors selecting
the space of physical states.

 However, we do not want to stick necessarily to the Dirac formalism, in particular it can be
interesting to study the intermediate step where particular roles are given to some of the sources and to study partial observables which do not commute with some of the $\pi_{(k)}.$
 Indeed, in the ${\mathfrak g}=sl(2,{\mathbb C})_{\mathbb R}$ case a Chern-Simons source is a free massive spining
 point particle evolving in deSitter space. In this case the weight $\chi$ is expressed in term of
 the mass and the spin of the particle. The formalism of Chern-Simons theory coupled to sources
 describes the particles coupled to deSitter gravity.
Dirac quantization program, which can
 be fully understood using the combinatorial quantization techniques, allow us to quantize and analyze only the
 quantum constants of motion of the system. This  leads inevitably to the frozen time problem.
 It is therefore important to design alternative quantization schemes in order to quantize larger class of
 observables which are not Dirac observables, for examples  partial observables involved in a conditional probability
 description of quantum gravity.
The hierarchy of Poisson algebras that we have introduced in this section has been designed for this purpose.

 We will therefore study in the next sections the quantization of the Poisson
 algebras $Hol_p(\mathfrak{G}),$ $ Hol_p(\bullet,\mathfrak{G}),$ $ Hol_p(S^2,\mathfrak{G}),$ 
 We will study as  well  their star structures, which implement the choice of real structure on $\mathfrak{G}$ when 
 $\mathfrak{g}$ is of rank $1$ or when $\mathfrak{g}$ is $sl(2,\mathbb{C})_{\mathbb R}.$

This will be important for the construction and analysis of unitary
representations of these algebras which will be the subject of a second article \cite{BRpreparation}.

The canonical analysis can be pursued for the case of sources coupled to Chern-Simons gauge theory. In the case where $G=SL(2,\mathbb{C})_{\mathbb R}$, the connection is a $sl(2,\mathbb{C})\oplus sl(2,\mathbb{C})$
connection with a left and right component denoted $A^{(l)}, A^{(r)}.$ The Chern-Simons action coupled to sources is written as:
\begin{equation}
S[A,M_{(1)},..,M_{(p)}] =\frac{1}{i}(S[A^{(l)},M^{(l)}_{(1)},..,M^{(l)}_{(p)}] -
S[A^{(r)},M^{(r)}_{(1)},..,M^{(r)}_{(p)}] )
\end{equation}
with coupling constant $\theta\in\mathbb{R}.$
The hamiltonian analysis has already been done and the final result is that the dynamical r-matrix in this case is simply:
\begin{eqnarray}
&&r(\varphi;-\tilde{\chi})=r^{(ll)\;\;\frac{\theta}{i}}_{12}(\varphi;-\tilde{\chi}^{(l)})+r^{(rr)\;\;-\frac{\theta}{i}}_{12}(\varphi;-\tilde{\chi}^{(r)})\\
&&=r^{(ll)\;\;\frac{\theta}{i}}_{12}(\varphi;-\tilde{\chi}^{(l)})-r^{(rr)\;\;\frac{\theta}{i}}_{21}(\varphi;-\tilde{\chi}^{(r)}).
\end{eqnarray}
In the following we will use the shortened notation $r^{(ll)(\pm)}_{12}(-\tilde{\chi}^{(l)})=r^{(ll)\;\;\frac{\theta}{i}}_{12}(0\pm;-\tilde{\chi}^{(l)})$
and $r^{(rr)(\pm)}_{12}(-\tilde{\chi}^{(r)})=r^{(rr)\;\;\frac{\theta}{i}}_{12}(0\pm;-\tilde{\chi}^{(r)}).$\\

 The corresponding Poisson algebra of dynamical boundary-boundary holonomies is 
 generated by elements $V^{(l)}_{\gamma_{(k)}},V^{(r)}_{\gamma_{(k)}},\tilde{\chi}_{(k)}^{(l)},\tilde{\chi}_{(k)}^{(r)}$ such that
\begin{eqnarray}
&&\{V^{(\sigma)}_{\gamma_{(k)}}, \tilde{\chi}_{(1)}^{(\epsilon)}\}=V^{(\sigma)}_{\gamma_{(k)}} h^{(\sigma)}\delta_{\sigma,\epsilon}\;\;,\;\;
\{V^{(\sigma)}_{\gamma_{(k)}}, \tilde{\chi}_{(m)}^{(\epsilon)}\}=-\delta_{k,m}h^{(\sigma)}
 V^{(\sigma)}_{\gamma_{(k)}}\delta_{\sigma,\epsilon},\;\;\\
 &&\{V^{(l)}_{\gamma_{(k)}1},V^{(r)}_{\gamma_{(m)}2}\}=0\\
&&\{V^{(l)}_{\gamma_{(k)}1},V^{(l)}_{\gamma_{(m)}2}\}=
V^{(l)}_{\gamma_{(k)1}}V^{(l)}_{\gamma_{(m)2} }r^{(ll)(-)}_{12}
(-\tilde{\chi}^{(l)}_{(1)})\;\;\text{with}\;\;k<m,
\label{poissongrav1}\\
&&\{V^{(r)}_{\gamma_{(k)}1},V^{(r)}_{\gamma_{(m)}2}\}=
V^{(r)}_{\gamma_{(k)1}}V^{(r)}_{\gamma_{(m)2} }r^{(rr)(+)}_{12}
(-\tilde{\chi}^{(r)}_{(1)})\;\;\text{with}\;\;k<m,
\label{poissongrav1a}\\
&&\{V^{(\sigma)}_{\gamma_{(k)} 1},V^{(\epsilon)}_{\gamma_{(k)} 2}\}=\delta_{\sigma,\epsilon}(V^{(\sigma)}_{\gamma_{(k)} 1}
V^{(\sigma)}_{\gamma_{(k)} 2} r^{(\sigma\sigma)(-)}_{12}(-\tilde{\chi}^{(\sigma)}_{(1)}) +\nonumber\\
&&\hskip 3cm+r^{(\sigma\sigma)(+)}_{12}(-\tilde{\chi}^{(\sigma)}_{(k)})V^{(\sigma)}_{\gamma_{(k)} 1}
V^{(\sigma)}_{\gamma_{(k)} 2}).\label{poissongrav2b}
\end{eqnarray}
\bigskip
Note that in this case we can define a star structure on $Hol_p(sl(2,{\mathbb C})\oplus sl(2,{\mathbb C}))$ as 
$(\star \otimes \star)(V^{(l)}_{\gamma_{(k)}})=V^{(r)\;-1}_{\gamma_{(k)}},
(\star \otimes \star)(\tilde{\chi}_{(k)}^{(l)})=-\tilde{\chi}_{(k)}^{(r)}.$

\section{Quantization of Chern-Simons sources}

We now proceed to the quantization of a free Chern-Simons source i.e to the quantization of the Poisson algebra ${\cal
S}(\mathfrak{G}).$

The quantization of dynamical quadratic Poisson brackets is a well studied subject and the quantization of 
the Poisson brackets of the free source gives:
\begin{eqnarray}
&&M_1 M_2=M_2 M_1 R_{12}(\hat{\mu})\label{quantumfreealgebra1}\\
&&[ \hat{\mu}_{\alpha_i}, M]=Mh_{\alpha_i},\label{quantumfreealgebra2}
\end{eqnarray}
with the following semiclassical behaviour
$\hat{\mu}\sim -\frac{\tilde{\chi}}{i \hbar}$ and $R_{12}(\hat{\mu})=1+i\hbar r_{12}(-\tilde{\chi})+o(\hbar).$
We will denote $\mu=-\frac{\chi}{i\hbar}.$

It is natural, due to the associativity of the operator algebra
(\ref{quantumfreealgebra1},\ref{quantumfreealgebra2}),  to require that the matrix $R(\mu)$ satisfies
the dynamical Yang-Baxter equation (\ref{qdybe}) with $R(\mu)\in U(\mathfrak{G})^{\otimes 2}.$

We define an  algebra denoted $\hat{\cal S}(\mathfrak{G})$, generated by the components of  $M\in U(\mathfrak{G})\otimes
 \hat{\cal S}(\mathfrak{G}) $ and components of $\hat{\mu} \in  {\mathfrak H}\otimes\hat{\cal S}(\mathfrak{G}),$ with relations:
\begin{eqnarray}
&&M_1M_2=( \Delta\otimes id )(M) F_{12}(\hat{\mu})\;\;,\label{freeCSsource1}\\
&&[ \hat{\mu}_{\alpha_i}, M]=M h_{\alpha_i}.\label{freeCSsource2}
\end{eqnarray}
Note that the relation (\ref{freeCSsource1}) implies the exchange relation (\ref{quantumfreealgebra1}).

$\hat{\mu}_{\alpha_i}$ are elements which generate gauge transformations. The algebra of gauge invariant elements, 
denoted $\hat{\cal S}(\mathfrak{G})^{\mathfrak H},$ is   the subalgebra of $\hat{\cal S}(\mathfrak{G})$  commuting
with components of $\hat{\mu}$.
It is generated by the components of 
\begin{equation}
P=\frac{i\hbar}{2}M\;(b(\hat{\mu})+c)\;M^{-1}\label{defofPfreequantum}
\end{equation}
where $b(\hat{\mu})=\sum_j (2\hat{\mu}_j+h_{\alpha_j}) \lambda^j$ and $c$ is the quadratic Casimir of $U(\mathfrak{G})$ normalized by $\Delta(c)=c_1+c_2+2 t_{12}.$

$P$ satisfy
\begin{eqnarray}
&&\Delta(P)=P_1+P_2,\label{DeltaonPquantique}\\
&&[P_1,M_2]=i\hbar t_{12}M_2\;\;,\;\;\;\;\;\; [P_1,P_2]=i\hbar [P_1,t_{12}]\;\;,\;\;\;
\;\;\;[P_1,\hat{\mu}_2]=0.\label{CommutatorofP}
\end{eqnarray}
As a result, we  obtain that $P=t^{ab}X_a\otimes \hat{P}_b$, with
$[\hat{P}_a,\hat{P}_b]=i\hbar f_{ab}^c \hat{P}_c,$  where $[X_a,X_b]=f_{ab}^c X_c.$ The algebra generated by the components of
 $P,$ i.e $(\hat{P}_a)$ is therefore isomorphic to $U(\mathfrak{G}).$
The proof of the results (\ref{DeltaonPquantique},\ref{CommutatorofP}) is contained in Appendix 4.

In a Dirac quantization process the study of unitary representations of this algebra is sufficient. However,  
in order to follow other quantization schemes, it is necessary to study representation of the algebra
$\hat{\cal S}(\mathfrak{G})$ itself.

In the case where $\mathfrak{G}=sl(2, \mathbb{C})$, we have the formula:
\begin{eqnarray}
&&F(\mu)=1-\frac{1}{\mu_0}e\otimes f\;\;\;\;\text{in the fundamental representation},
\end{eqnarray}
where we have denoted by $\mu_0$ the quantity $\mu_{\alpha}$ where $\alpha$ is the positive root of $sl(2).$

As a result $R(\mu)=1-\frac{1}{\mu_0}e\wedge f-\frac{1}{\mu_0^2}fe\otimes ef$ in the fundamental
representation.

The equations (\ref{quantumfreealgebra1},\ref{quantumfreealgebra2}) defining
$\hat{\cal S}(\mathfrak{G} )$ can be written:
\begin{eqnarray}
&&M_1 {} M_2 {} =M_2 {} M_1 {} R_{12}(\hat{\mu}_0 {} )\nonumber\\
&&\hat{\mu}_0 M= M (\hat{\mu}_0+h).\label{comrelfree}
\end{eqnarray}
Denoting  $g=({\buildrel {\frac{1}{2}} \over \pi} \otimes id)(M)=\left( \begin{array}{cc} a {}  & b {} \\c {}  & d {}  \end{array}
\right),$ the commutation relations (\ref{comrelfree}) can be rewritten as:

\begin{eqnarray}
&&\hat{\mu}_0 {}  b {} =b {}  (\hat{\mu}_0 {} -1),\;\;\;\;\;\;\hat{\mu}_0 {}  d {} =
d {}  (\hat{\mu}_0 {} -1),\;\;\;\;\;\;\hat{\mu}_0 {}  a {} =a {}  (\hat{\mu}_0 {} +1),\;\;\;\;\;\;
\hat{\mu}_0 {}  c {} =c {}  (\hat{\mu}_0 {} +1)\nonumber\\
&&a {} c {} =c {} a {} , \;\;\;\;\;\;\;\; b {} d {} =d {} b {} ,\;\;\;\;\;\;\;\;
 a {} b {}  (1-\frac{1}{\hat{\mu}_0 {} })=b {} a {} , \;\;\;\;\;\;\;\;
c {} d {}  (1-\frac{1}{\hat{\mu}_0 {} })=d {} c {} \nonumber\\
&&a {} d {} -d {} a {} =c {} b {} \frac{1}{\hat{\mu}_0 {} },\;\;\;\;\;\;\;\;\;\;
 c {} b {} -b {} c {} =a {} d {} \frac{1}{\hat{\mu}_0 {} }.
\end{eqnarray}
The quantum determinant $a {} d {} -c {} b {} $  is a central element which is fixed to 
\begin{eqnarray}
a {} d {} -c {} b {} =1\label{quantumdetfree}
\end{eqnarray}
from the fusion equation (\ref{freeCSsource1}).

 In the Dirac scheme, the first class constraints $\Pi {} =\hat{\mu}_0 {} -\mu_0$ are
 promoted to constraint operators. The algebra of quantum observables $\hat{\cal S}(\mathfrak{G}){}^{\mathfrak{H}}$ is the 
 subalgebra of operators commuting with $\Pi {} $.
As in the classical case, the algebra of quantum observables  $\hat{\cal S}(\mathfrak{G}){}^{\mathfrak{H}}$ is then generated by
momentum variables $P$ and is isomorphic to the algebra ${U}(sl(2,\mathbb{C}))$ using
 \begin{eqnarray}
e=-c {} \hat{\mu}_0 {} d {},\;\;\;f=a {} \hat{\mu}_0 {} b {},\;\;\; h {} =-
(a {} \hat{\mu}_0 {} d {} +c {} \hat{\mu}_0 {} b {} ).\label{efh}
\end{eqnarray}
The center of $ \hat{\cal S}(\mathfrak{G}){}^{\mathfrak{H}}$ is generated by the quadratic Casimir element which has a very simple
expression in term of $\hat{\mu}_0:$
$C {} =h^2+2(e {} f {} +f {} e {}) =\hat{\mu}_0{}^2-1.$

Along the lines of the classical treatment, we introduce the following linear automorphisms of $\hat{\cal S}(\mathfrak{G})$:
\begin{eqnarray}
&&\sigma_1(a)\!=\!a\;\;\;\;\sigma_1(b)\!=\!-b\;\;\;\;\sigma_1(c)\!=\!-c\;\;\;\;\sigma_1(d)\!=\!d\;\;\;\;\sigma_1(\hat{\mu}_0)\!=\!\hat{\mu}_0\\
&&\sigma_2(a)\!=\!\frac{\hat{\mu}_0}{\hat{\mu}_0+1}d\;\;\;\;\;\sigma_2(b)\!=\!-c\;\;\;\;\;\sigma_2(c)\!=\!-\frac{\hat{\mu}_0}{\hat{\mu}_0+1}b\;\;\;\;\;
\sigma_2(d)\!=\!a\;\;\;\;\;\sigma_2(\hat{\mu}_0)\!=\!-\hat{\mu}_0.\nonumber\\&&
\end{eqnarray}
In order to identify hermitian operators associated to real functions on the classical phase space, we
have to define a star structure (involutive antilinear antimorphism) on the complex 
algebra defined by the relations (\ref{freeCSsource1},\ref{freeCSsource2}) quantizing the function algebra according to its classical counterpart defined in previous section.
In the case of first rank algebras, there are only three cases of interest:
\begin{eqnarray}
&& \hat{\cal S}({su(2)}):a^\star=d\;\;\;\;b^\star=-c\;\;\;\;\hat{\mu}_0^\star=\hat{\mu}_0,\;\;\;\;\\
&& \hat{\cal S}({su(1,1)}):{\overline{\star}}= \sigma_1 \circ \star, \;\;\;\;\\
&& \hat{\cal S}({sl(2,\mathbb{R})}):{\underline{\star}}=\sigma_2 \circ \star .\nonumber\\&&
\end{eqnarray}
As a remark we have the properties $\sigma_i \circ \star=\star \circ \sigma^{-1}_i$ necessary for these stars to be involutive.
Using definitions (\ref{efh}) we obtain
\begin{eqnarray}
&&su(2):e^{\star}=f,\;\;\;\;f^{\star}=e,\;\;\;\; h^{\star}=h\\
&&su(1,1):e^{\overline{\star}}=-f,\;\;\;\;f^{\overline{\star}}=-e,\;\;\;\; h^{\overline{\star}}=h\\
&&sl(2,\mathbb{R}):e^{\underline{\star}}=-e,\;\;\;\;f^{\underline{\star}}=-f,\;\;\;\; h^{\underline{\star}}=-h.
\end{eqnarray}

In the $sl(2,\mathbb{C})_{\mathbb{R}}$ case, the algebra of interest is 
$\hat{\cal S}(sl(2,\mathbb{C})_{\mathbb{R}})=\hat{\cal S}(sl(2,\mathbb{C}))^{\otimes 2}$ equipped with one of the following star structures:
\begin{eqnarray}
&&a^{(l)}{}^\bigstar\!\!=d^{(r)}\;\;\;\;b^{(l)}{}^\bigstar\!\!=-c^{(r)}\;\;\;\;c^{(l)}{}^\bigstar\!\!=- b^{(r)}\;\;\;\;\;
d^{(l)}{}^\bigstar\!\!=a^{(r)}\;\;\;\;\hat{\mu}^{(l)}{}^{\bigstar}\!\!=\hat{\mu}^{(r)},\nonumber\\
&&{}^{\overline{\bigstar}}=(id \otimes \sigma_1) \circ {}^{\bigstar}\;\;\;\;\;\;\;\;{}^{\underline{\bigstar}}=(id \otimes \sigma_2) \circ {}^{\bigstar}.
\end{eqnarray}
In each of these cases, it can be shown that the star Lie algebra generated by the components of $P$  are all isomorphic to 
$sl(2,{\mathbb C})_{\mathbb R}.$

\section{Quantization of  Chern-Simons theory coupled to Sources}
\label{sectionquantum}

In this section we define and study the quantization of the Poisson algebras defined in section \ref{sectionclassic}.

The quantization of the Poisson algebra $Hol_p(\mathfrak{G})$ is the algebra denoted
$\widehat{Hol}_p(\mathfrak{G})$ and defined as:
\begin{eqnarray}
&&[\hat{\mu}_{(1)\alpha_i},V_{\gamma_{(k)}}]=V_{\gamma_{(k)}} h_{\alpha_i}\;\;,\;\;\nonumber
[\hat{\mu}_{(l)\alpha_i},V_{\gamma_{(k)}}]=-h_{\alpha_i}
 V_{\gamma_{(k)}}\delta_{kl},\;\;\\
&&V_{\gamma_{(k)} 1}V_{\gamma_{(l)} 2}=V_{\gamma_{(l)} 2}V_{\gamma_{(k)} 1}
 R^{\theta (-)}_{12}
(\hat{\mu}_{(1)})\;\;\text{with}\;\;k<l,
\label{quantumgrav1}
\end{eqnarray}
where $R^{\theta}(\hat{\mu})$ is a solution of dynamical Yang-Baxter equation, with
 $R^{\theta(+)}(\hat{\mu})=R^{\theta}(\hat{\mu}),$ $
R^{\theta(-)}(\hat{\mu})=R^{\theta}_{21}(\hat{\mu})^{-1},$ and  the following semiclassical behaviour
$\hat{\mu}_{(k)}\sim -\frac{\tilde{\chi}_{(k)}}{i \hbar}$ and
 $R_{12}(\hat{\mu}_{(k)})=1+i\hbar r_{12}^{\theta}(-\tilde{\chi}_{(k)})+o(\hbar)$ and define
  $ q=exp(\frac{i\hbar}{4\theta}).$

We will denote $\mu_{(k)}=-\frac{\chi_{(k)}}{i \hbar}.$

The quantization of (\ref{poissonV2}) leads to
\begin{eqnarray}
&& R^{\theta (\epsilon)}_{21}(\hat{\mu}_{(k)})
V_{\gamma_{(k)} 1}
V_{\gamma_{(k)} 2}
=V_{\gamma_{(k)} 2}
V_{\gamma_{(k)} 1}
R^{\theta (\epsilon)}_{12}(\hat{\mu}_{(1)}),
\end{eqnarray}
which is implied by
\begin{eqnarray}
&&
V_{\gamma_{(k)} 1}
V_{\gamma_{(k)} 2}
=F^{\theta\; -1}_{21}(\hat{\mu}_{(k)})R_{12}^{\theta}(\Delta \otimes id)(V_{\gamma_{(k)}})
F^{\theta}_{12}(\hat{\mu}_{(1)}).\label{quantumgrav2}
\end{eqnarray}
The set of relations (\ref{quantumgrav1},\ref{quantumgrav2})
defines the quantum dynamical algebra $\widehat{Hol}_p(\mathfrak{G}).$

According to the classical case, we will have to implement the relations imposed
by the flatness of the connection. In order to achieve this, we will quantize
the Poisson algebra  ${H}ol_p(\bullet,\mathfrak{G}).$
As in the classical case, we have to choose a $R-$matrix denoted $\breve{R}_{12}^{\theta}$ solution 
of quantum Yang-Baxter equation, quantizing the classical $r-$matrix $\breve{r}_{12}^{\theta}.$ These 
$R-$matrices are of the form $\breve{R}_{12}^{\theta}=J^{-1}_{21}R_{12}^{\theta}J_{12}$ where $J$ is any
2-cocycle. Different choices will lead
to the same results after having moded out by the quantum flatness condition. Hence, in absence
 of explicit mention, we will choose $J=1\otimes 1$ (however in the $sl(2,\mathbb{C})$ case we will choose a non
 trivial $J$).

We then define the algebra $\widehat{Hol}_p(\bullet,\mathfrak{G})$ by the relations:
\begin{eqnarray}
&&[\hat{\mu}_{(k)\alpha_i},U_{(l)}]=\delta_{kl}U_{(k)} h_{\alpha_i}\;\;,\;\;\\
&&U_{(k) 1}U_{(k) 2}=J_{21}^{-1}R_{12}^{\theta}(\Delta \otimes id)(U_{(k)})
F^{\theta}_{12}(\hat{\mu}_{(k)}).\label{fusionUquantique}\\
&&\breve{R}^{\theta}_{21}U_{(k) 1}U_{(l) 2}=U_{(l) 2}U_{(k) 1}\;\;\;\;k<l.
\end{eqnarray}
Note that relation (\ref{fusionUquantique}) implies the exchange relations:
\begin{equation}
\breve{R}_{12}^{\theta (\pm)-1}U_{(k) 1}U_{(k) 2}=U_{(k) 2}U_{(k) 1}
R^{\theta (\mp)}_{12}(\hat{\mu}_{(k)}).
\end{equation}

Let us introduce the elements ${\cal P}_{(k)}$ which are the q-analogue  of the element (\ref{defofPfreequantum})
\begin{equation}
{\cal P}_{(k)}=U_{(k)} v^{-1} B(\hat{\mu}_{(k)}) U_{(k)}^{-1}\;\;,\;\;k=1,...,p.
\end{equation}

These elements commute with the action of the Cartan elements and satisfy the following relations:
\begin{eqnarray}
&&[\hat{\mu}_{(k)\; \alpha_i},{\cal P}_{(l)}]=0,\;\;\forall k,l\\
&&{\cal P}_{(k)\;2}\breve{R}_{12}^{\theta\;-1}{\cal P}_{(k)\;1}\breve{R}^{\theta}_{12}= J_{12}^{-1}(\Delta \otimes id)({\cal P}_{(k)})J_{12}\label{deltasurlescalP}\\
&&\breve{R}^{\theta}_{21}{\cal P}_{(k)\;1}\breve{R}^{\theta \; -1}_{21}{\cal P}_{(l)\;2}=
{\cal P}_{(l)\;2}\breve{R}^{\theta}_{21}{\cal P}_{(k)\;1}\breve{R}^{\theta\;-1}_{21},\\
&&\breve{R}^{\theta}_{21}U_{(k)\;1}{\cal P}_{(k)\;2}={\cal P}_{(k)\;2} \breve{R}^{\theta\;-1}_{12} U_{(k)\;1}\label{UcalP}\\
&&\breve{R}^{\theta\;-1}_{12}U_{(l)\;1}{\cal P}_{(k)\;2}={\cal P}_{(k)\;2} \breve{R}^{\theta\;-1}_{12} U_{(l)\;1}\\
&&\breve{R}^{\theta}_{21}U_{(k)\;1}{\cal P}_{(l)\;2}={\cal P}_{(l)\;2} \breve{R}^{\theta}_{21} U_{(k)\;1}
\;\;\forall k<l.
\end{eqnarray}
These identities are straightforwardly derived from basic definitions, and we give, as an example, the proof of (\ref{deltasurlescalP}) and (\ref{UcalP}) in the appendix 4.

A central result is the following factorization theorem:
$$\widehat{{H}ol}_p(\bullet,\mathfrak{G})\;\; \text{is isomorphic to} \;\;
\widehat{{H}ol}_1(\bullet,\mathfrak{G})^{\otimes p}.$$
We will now prove this proposition.
$\widehat{{H}ol}_1(\bullet,\mathfrak{G})$ is the algebra defined by
\begin{eqnarray}
&&[\hat{\mu}_{\alpha_i},W]=W h_{\alpha_i}\;\;,\;\;\nonumber\\
&&W_{1}W_{2}=J_{21}^{-1}R_{12}^{\theta}(\Delta \otimes id)(W)
F^{\theta}_{12}(\hat{\mu}),\label{definitionW}
\end{eqnarray} where we have denoted $W=U_{(1)}, \hat{\mu}=\hat{\mu}_{(1)}.$

This algebra contains $U_q({\mathfrak G})$ as a subalgebra. Indeed if we define
$L=W v^{-1} B(\hat{\mu})W^{-1}$ we verify
\begin{eqnarray}
&&L_{2}\breve{R}_{12}^{\theta\;-1}L_{1}\breve{R}^{\theta}_{12}= J_{12}^{-1}(\Delta \otimes id)(L)J_{12}.
\end{eqnarray}
As usual we introduce the Gauss factorization $L=L^{(+)\;-1}L^{(-)}$ with 
\begin{eqnarray}
L_1^{(\pm)}L^{(\pm)}_2&=&J_{12}^{-1}\Delta(L^{(\pm)})J_{12}\\
\breve{R}^{\theta}_{12}L_1^{(+)}L^{(-)}_2&=&L_2^{(-)}L_1^{(+)}\breve{R}^{\theta}_{12}.
\end{eqnarray}
Moreover $U_q({\mathfrak G})$ acts on $\widehat{{H}ol}_1(\bullet,\mathfrak{G})$ by adjoint action as follows
\begin{eqnarray}
L_2^{(\pm)\;-1}W_1 L_2^{(\pm)}=\breve{R}^{\theta\;(\pm)}_{21}W_1\;\;\;\;\;\;
L^{(\pm)\;-1} \hat{\mu} L^{(\pm)}=\hat{\mu}.
\end{eqnarray}

Let us now consider the algebra $\widehat{{H}ol}_1(\bullet,\mathfrak{G})^{\otimes p}$ as
being generated by $W_{(k)},\hat{\mu}_{(k)}$
associated to $p$ commuting copies of $\widehat{{H}ol}_1(\bullet,\mathfrak{G})$
(we will also denote $L^{(\pm)}_{(k)}$ the corresponding elements).
We introduce one more element by
\begin{eqnarray}
&&\Gamma^{(\pm)}_{(k)}=\prod_{j=k}^p L^{(\pm)}_{(j)}\;\;\;\;\forall k=1,...,p\;\;\;\;\;\;\;\Gamma^{(\pm)}_{(p+1)}=1.
\end{eqnarray}

It is easy to show that  following map
\begin{eqnarray}
\kappa:\widehat{{H}ol}_p(\bullet,\mathfrak{G}) &\longrightarrow& \widehat{{H}ol}_1(\bullet,\mathfrak{G})^{\otimes p}\nonumber\\
(U_{(k)},\hat{\mu}_{(k)}) &\mapsto& (\Gamma^{(-)\;-1}_{(k+1)} W_{(k)},\hat{\mu}_{(k)})
\end{eqnarray}
is an isomorphism of algebra.

In order to implement the conditions  of the flatness of the connection it is useful
to remark  that
\begin{eqnarray}
\kappa ( {\cal P}_{(k)} )&=&\Gamma^{(-)\;-1}_{(k+1)}L_{(k)} \Gamma^{(-)}_{(k+1)},\\
\kappa ( \prod_{j=p}^{1}{\cal P}_{(j)} )&=&\Gamma^{(+)\;-1}_{(1)}\Gamma^{(-)}_{(1)}.
\end{eqnarray}

The subalgebra of elements of $\widehat{{H}ol}_p(\bullet,\mathfrak{G})$ commuting with  
$\prod_{j=p}^{1}{\cal P}_{(j)}$ will be denoted
$\widehat{{H}ol}_p(\bullet, \mathfrak{G})^{U_q(\mathfrak{G})}\simeq \widehat{{H}ol}_p(\mathfrak{G}).$

 Note that if  ${\cal H }$ is a $\widehat{{H}ol}_1(\bullet,\mathfrak{G})$ module,
using the explicit isomorphism between $\widehat{{H}ol}_p(\bullet,\mathfrak{G})$ and
$\widehat{{H}ol}_1(\bullet,\mathfrak{G})^{\otimes p}$, we obtain that a representation of 
$\widehat{{H}ol}_p(\bullet,\mathfrak{G})$ is provided by the module ${\cal H }^{\otimes p}.$
Because $U_q(\mathfrak{G})$ acts on  ${\cal H }^{\otimes p}$ via  $\Delta^{(p)}$, this action corresponds 
to the action of $\Gamma^{(\pm)}_{(1)}.$
Therefore a representation   of $\widehat{{H}ol}_p(\bullet,\mathfrak{G}))^{U_q(\mathfrak{G})}$ 
implementing the condition $\prod_{j=p}^{1}{\cal P}_{(j)}=1$ consists in the module 
$({\cal H }^{\otimes p}){}^{U_q(\mathfrak{G})}$ i.e the vector space of invariant vectors under the action of
$U_q(\mathfrak{\mathfrak{G}}) .$

We will construct an explicit unitary representation of $\widehat{{H}ol}_p(\mathfrak{G})$ when 
$\mathfrak{g}=sl(2,\mathbb{R})$ or $\mathfrak{g}=sl(2,\mathbb{C})_{\mathbb{R}}$ in \cite{BRpreparation}.

We will now provide, in the case where $\mathfrak{G}=sl(2,\mathbb{C}),$ an explicit analysis of the  
structure of
 $\widehat{{H}ol}_1(\bullet,\mathfrak{G}).$  Let us denote $\hat{x}=q^{\hat{\mu}_0},$ and
$({\buildrel {\frac{1}{2}} \over \pi} \otimes id)(W)=\left( \begin{array}{cc} a {}  & b {} \\c {}  & d {}  \end{array}
\right).$ 
In the fundamental representation we have:
\begin{equation}
F^{\theta}_{12}(x)=1-(e\otimes f)\frac{q-q^{-1}}{1-x^{-2}}.
\end{equation}
As a result the exchange relations implied by (\ref{definitionW}) can be explicitely recast as:
\begin{eqnarray}
&&\hat{x}a=qa\hat{x},\;\;\;\;\hat{x}c=qc\hat{x},\;\;\;\;\hat{x}b=q^{-1}b\hat{x},\;\;\;\;\hat{x}d=q^{-1}d\hat{x},\nonumber\\
&&ac=qca, \;\;\;\;bd=qdb, \;\;\;\;ab (q^{-1}-\frac{q-q^{-1}}{\hat{x}^2-1})=ba, \;\;\;\;
cd (q^{-1}-\frac{q-q^{-1}}{\hat{x}^2-1})=dc,\nonumber\\
&&ad-da=cb\frac{q-q^{-1}}{1-\hat{x}^{-2}},\;\;\;\;\;\;cb-bc=ad\frac{q-q^{-1}}{\hat{x}^2-1}..\label{commquantique}
\end{eqnarray}
$ad-qcb$ is a central element which can be computed using the fusion relation (\ref{definitionW})
\begin{eqnarray}
&&ad-qcb=q^{-1/2}.\label{qdet}
\end{eqnarray}
The subalgebra $U_q(sl(2))$ is expressed as:
\begin{equation}
({\buildrel {\frac{1}{2}} \over \pi} \otimes id)(L)=\left( \begin{array}{cc} (q-q^{-1})^2 qfe+q^{-h} {}  & -(q-q^{-1})q^2q^h f {} \\-q(q-q^{-1})e {}  & q^h {}  \end{array}
\right)=({\buildrel {\frac{1}{2}} \over \pi} \otimes id)(Wv^{-1}B(\hat{\mu})W^{-1}),
\end{equation}
which can be written 
\begin{eqnarray}
&& e=-cd \frac{q^{\frac{3}{2}}(q^{-1}\hat{x}-q\hat{x}^{-1})}{q-q^{-1}} ,\;\;\;\;
q^h \!f=ab\frac{q^{-\frac{1}{2}}(\hat{x}q^{-1}-\hat{x}^{-1}q)}{q-q^{-1}}, \nonumber\\
&&q^h=q^{\frac{1}{2}}(a\hat{x}^{-1}d-qc\hat{x}b).
\end{eqnarray}
The Casimir element $C=(q-q^{-1})^2fe+q q^h+q^{-1}q^{-h}$ is given by
$C=\hat{x}+\hat{x}^{-1}.$

We will now study when  $\mathfrak{G}=sl(2,\mathbb{C}),$ the star structures that can de defined on  
$\widehat{{H}ol}_1(\bullet,\mathfrak{G}).$ In the case where $\mathfrak{g}$ is a simple Lie algebra, 
our quantization procedure leads to the expression $q=exp(\frac{i\hbar}{4\theta}),$ therefore  
  $q$ has to be unimodular.

We will not dwell on  the case where $\mathfrak{g}=su(2).$ In that case numerous technical problem arise, in this
formalism as well as others (need of weak quasi-Hopf algebras at the start to ensure truncation of the spectrum
for $q$ root of unit, etc..) and we do not want to adress them in this work.

We will only give here the $\star$ structure associated to $\mathfrak{g}=sl(2,\mathbb{R}).$ 
Let us now consider the star algebra 
$(\widehat{{H}ol}_1(\bullet,sl(2,\mathbb{C})),{\underline{\star}})$ associated to the
 $sl(2,{\mathbb R})$ case. The star structure, in this case, is defined by:
\begin{eqnarray}
\vert  q \vert=1
&&a^{\underline{\star}}=q^{2}a\;\;\;\;b^{\underline{\star}}=q\frac{x^2-1}{q^2x^2-1}b\;\;\;\;c^{\underline{\star}}=qc\;\;\;\;d^{\underline{\star}}=\frac{x^2-1}{q^2x^2-1}d\;\;\;\;\hat{x}^{\underline{\star}}=x.
\end{eqnarray}
As a consequence, the star structure on the subalgebra $U_q(sl(2,{\mathbb R}))$ is given by
\begin{eqnarray}
&&e^{\underline{\star}}=-q^{-1}e\;\;\;\;\;\;(q^h)^{\underline{\star}}=q^h\;\;\;\;\;\;f^{\underline{\star}}=-qf.
\end{eqnarray}

In the case where $\mathfrak{g}=sl(2,{\mathbb C})_{\mathbb R}$ our choice of classical invariant bilinear form 
imposes   $q$ to be fixed to the real value $q=exp(\frac{\hbar}{4\theta}).$ A detailed study of 
$U_q(sl(2,{\mathbb C})_{\mathbb R})$ has been done in \cite{Harmonic}. In this work we have chosen
$ U_q(sl(2,{\mathbb C})_{\mathbb R})$ being the quantum double of $U_q(su(2))$. This choice implements 
naturally a quantum analog of Iwasawa decomposition allowing to develop  an harmonic analysis on 
$U_q(sl(2,{\mathbb C})_{\mathbb R}).$
We have $U_q(sl(2,{\mathbb C})_{\mathbb R})=U_q(sl(2))^{(l)}\otimes U_q(sl(2))^{(r)}$ as an algebra. The isomorphism of
coalgebra is true up to a twist $J=(R^{\theta\;}_{12})^{(rl)}.$  As a result 
the $R$ matrix of 
the quantum double is twist equivalent to $(R^{\theta}_{12})^{(ll)}(R^{\theta\;(-)}_{12})^{(rr)}$ 
with the twist $J=(R^{\theta\;}_{12})^{(rl)}.$

As a result we define
$\widehat{{H}ol}_1(\bullet,(sl(2,{\mathbb C})_{\mathbb R})^{\mathbb C})$ to be the   algebra:
\begin{eqnarray}
&&[\hat{\mu}^{(\sigma)},W^{(\epsilon)}]=W^{(\sigma)} h^{(\sigma)}\delta_{\sigma,\epsilon}\;\;,\;\;\\
&&W^{(l)}_{1}W^{(l)}_{2}=(R^{\theta}_{12})^{(ll)}(\Delta \otimes id)(W^{(l)})
F^{\theta}_{12}(\hat{\mu}^{(l)})\\
&&W^{(r)}_{1}W^{(r)}_{2}=(R^{\theta\;(-)}_{12})^{(rr)}(\Delta \otimes id)(W^{(r)})
F^{\theta}_{12}(\hat{\mu}^{(r)})\\
&&(R^{\theta}_{21})^{(lr)}W^{(l)}_{1}W^{(r)}_{2}=W^{(r)}_{2}W^{(l)}_{1}.\label{exchangerelation}
\end{eqnarray}
As a result this  algebra is generated by the elements 

$\hat{x}^{(l)}=q^{\hat{\mu}^{l}},
\hat{x}^{(r)}=q^{\hat{\mu}^{r}}$ as well as the matrix elements of
\begin{eqnarray}
&&({\buildrel {(\frac{1}{2},0)} \over \pi} \otimes id)(W^{(l)})=\left( \begin{array}{cc} a^{(l)} {}  & b^{(l)} {} \\c^{(l)} {}  & d^{(l)} {}  \end{array}
\right),\;\;\;\;\;\;({\buildrel {(0,\frac{1}{2})} \over \pi} \otimes id)(W^{(r)})=\left( \begin{array}{cc} a^{(r)} {}  & b^{(r)} {} \\c^{(r)} {}  & d^{(r)} {}  \end{array}
\right).
\end{eqnarray}
The precise commutation relations are 
\begin{enumerate}
\item the left variables satisfy the relations (\ref{commquantique}) with $ a^{(l)}d^{(l)}-qc^{(l)}b^{(l)}=q^{-1/2}$
\item the right variables satisfy the relations (\ref{commquantique}) with $ a^{(r)}d^{(r)}-qc^{(r)}b^{(r)}=q^{5/2}$
\item the left and right variables satisfy the exchange relation (\ref{exchangerelation}).
\end{enumerate}

The subalgebra $U_q(sl(2,{\mathbb C}))\otimes U_q(sl(2,{\mathbb C}))$ is defined as:
\begin{eqnarray}
&& e^{(l)}=-\frac{q^{\frac{3}{2}}}{q-q^{-1}}c^{(l)}d^{(l)} (q^{-1}\hat{x}^{(l)}-q\hat{x}^{(l)\;-1}) ,\;\;\;\;
q^{h^{(l)}}\!\! f^{(l)}=a^{(l)}b^{(l)}\frac{q^{-\frac{1}{2}}(\hat{x}^{(l)}q^{-1}\!\!  -\! \hat{x}^{(l)\;-1}q)}{q-q^{-1}}, \nonumber\\
&&q^{h^{(l)}}=q^{\frac{1}{2}}(a^{(l)}\hat{x}^{(l)-1}d^{(l)}\!\!  -\! qc^{(l)}\hat{x}^{(l)}b^{(l)}),\;\;\;\;\;\;\;\; q^{h^{(r)}}=q^{-3}q^{\frac{1}{2}}(a^{(r)}\hat{x}^{(r)-1}d^{(r)}\!\!  -\! qc^{(r)}\hat{x}^{(r)}b^{(r)}),\nonumber\\
&& e^{(r)}=-q^{-3}c^{(r)}d^{(r)}\frac{q^{\frac{3}{2}}(q^{-1}\hat{x}^{(r)}\!\! -\!q\hat{x}^{(r)\;-1})}{q-q^{-1}}  ,\;\;\;\;
q^{h^{(r)}}\!\! f^{(r)}=q^{-3}a^{(r)}b^{(r)}\frac{q^{-\frac{1}{2}}(\hat{x}^{(r)}q^{-1}\!\!-\!\hat{x}^{(r)-1}q)}{q-q^{-1}}.\nonumber\\
&&
\end{eqnarray}

The star structure on $\widehat{{H}ol}_1(\bullet,(sl(2,{\mathbb C})_{\mathbb R})^{\mathbb C})$ is defined using
\begin{eqnarray}
a^{(l)\bigstar}\!=\!q^{-3/2}d^{(r)},\;\;b^{(l)\bigstar}\!=\!-q^{-1/2}c^{(r)},\;\;
c^{(l)\bigstar}\!=\!-q^{-5/2}b^{(r)},\;\;d^{(l)\bigstar}\!=\!q^{-3/2}a^{(r)},\;\; x^{(l)\bigstar}\!=\!x^{(r)}
\end{eqnarray}
Proving that this star is an antilinear antimorphism of $\widehat{{H}ol}_1(\bullet,sl(2,{\mathbb C})_{\mathbb
R}^{\mathbb C})$ is done by a direct verification of the commutation relations.
With this star we have, in particular,
\begin{eqnarray}
&&e^{(l)\bigstar}=q q^{h^{(r)}}\!\!f^{(r)}\;\;\;\;\;\;\;\;\;\;f^{(l)\bigstar}=q^{-1}e^{(r)}q^{-h^{(r)}}\;\;\;\;\;\;\;\;\;\;
(q^{h^{(l)}})^\bigstar=q^{h^{(r)}}.
\end{eqnarray}

Remark.
We end this section of the definition and study of quantization of these dynamical algebras by  looking at  the subalgebra of
$\widehat{Hol}_{p}(\mathfrak{G})$ generated by
$${\cal M}_{(k)}= V^{-1}_{\gamma_{(k)}}v^{-1}B(\hat{\mu}_{(k)})V_{\gamma_{(k)}}, k\geq 2$$ and 
$\hat{\mu}_{(1)}$.
We call this algebra the {\it quantum algebra of dynamical monodromies originating from the first source,} and we denote it 
 $\widehat{Mon}_p(\mathfrak{G}).$ 
The commutation relations satisfied by these elements are:
\begin{eqnarray}
&&[\hat{\mu}_{(1)\; \alpha_i},{\cal M}_{(k)}]=[{\cal M}_{(k)},h_{\alpha_i}],\label{monod1}\\
&&R_{21}(\hat{\mu}_{(1)}){\cal M}_{(k)\;1}R^{-1}_{21}(\hat{\mu}_{(1)}){\cal M}_{(l)\;2}=
{\cal M}_{(l)\;2}R_{21}(\hat{\mu}_{(1)}){\cal M}_{(k)\;1}R^{-1}_{21}(\hat{\mu}_{(1)}),\;\;k\!<\!l,\label{monod2}\\
&&{\cal M}_{(k)\;2}R^{-1}_{12}(\hat{\mu}_{(1)}){\cal M}_{(k)\;1}R_{12}(\hat{\mu}_{(1)})=
F^{-1}_{12}(\hat{\mu}_{(1)}) \; ( \Delta \otimes id)({\cal M}_{(k)}) \; F_{12}(\hat{\mu}_{(1)}).  \label{monod3}
\end{eqnarray}
As a result the algebra generated by $\hat{\mu}_{(1)}$ and ${\cal M}_{(k)}$ for  k fixed is a dynamical 
reflection algebra which has been recently studied in \cite{KM}.

As a remark, we note that   the equation (\ref{monod1}) and the linear equation  (\ref{ABRR2}) can be equivalently written as:
\begin{eqnarray}
&&(vB(\mu_{(1)}))^{-1}_2   R^{-1}_{12}(\hat{\mu}_{(1)})  {\cal M}_{(k)\;1}  R_{12}(\hat{\mu}_{(1)}) =
R_{21}(\hat{\mu}_{(1)})  {\cal M}_{(k)\;1}  R^{-1}_{21}(\hat{\mu}_{(1)})   (vB(\mu_{(1)}))^{-1}_2\label{monod5}\\
&&(vB(\mu_{(1)}))^{-1}_2  R^{-1}_{12}(\hat{\mu}_{(1)})  (vB(\mu_{(1)}))^{-1}_1    R_{12}(\hat{\mu}_{(1)})=
F^{-1}_{12}(\hat{\mu}_{(1)}) \; ( \Delta \otimes id)(  (vB(\mu_{(1)}))^{-1}   ) \; F_{12}(\hat{\mu}_{(1)}).  \nonumber
\end{eqnarray}

\section{Conclusion}

In this work, we have defined and studied algebras associated to the quantization of Chern-Simons theory with sources. We have emphasized the example $G=SL(2,{\mathbb R}) \times SL(2,{\mathbb R})$ and $G=SL(2,{\mathbb C})_{\mathbb R}$ having in mind potential applications to Lorentzian quantum gravity in $2+1$ dimensions with $\Lambda <0$ or $\Lambda >0.$ These algebras are bigger than the algebra of constants of motion and include partial observables depending on the parametrization of the world-line of the sources. These algebras are nice and new examples of dynamical exchange algebras.\\
 Our work can be extended in different directions that we are now exploring.\\
 \begin{enumerate}
 \item A direct continuation of our work is the representation side of the present work \cite{BRpreparation}. As explained in section \ref{}, the structure of $\widehat{Hol}_p(\bullet,G)$ is reduced to the study of $\widehat{Hol}_1(\bullet,G).$ We can construct therefore unitary representation of this algebra allowing to obtain unitary irreducible representation of $\widehat{Hol}_p(\bullet,G)$ and of its different quotients introduced in our work.
 \item {}From the mathematical perspective, we have pointed numerous links between these algebras. Our work can give new insights in the theory of dynamical quantum groups. In particular the study of the quantum algebra of dynamical monodromies
 will give new results in the explicit construction of the dynamical coboundary in the $sl(n)$ case.
\item The application of the present work to the study of physical questions that can be adressed in $2+1$ quantum gravity has to be developped and is currently in progress.
 \end{enumerate}

\bibliographystyle{unsrt}

\begin{thebibliography}{10}



\bibitem{BN}
E.~Buffenoir, K.~Noui, ``Unfashionable observations about three dimensional gravity''{\tt [arXiv: gr-qc/ 0305079  ] }.

\bibitem{ES} P.Etingof, O.Schiffmann,
``Lectures on the dynamical Yang-Baxter equations'',
{\tt [arXiv: math.QA/9908064]}.

\bibitem{AM} A.Alekseev, A.Malkin,``
Symplectic structure of the moduli space of flat connections on a Riemann surface'', Commun.Math.Phys. 169 (1995) 99 
{\tt [arXiv: hep-th/9312004]}.

\bibitem{Al}
A.Alekseev,
``Integrability in the Hamiltonian Chern-Simons Theory,''
st.Petersburg.Math.J, Vol.6 (1995),No.2,
{\tt [arXiv: hep-th/9311074]} 

\bibitem{AS}
A.Alekseev, V.Schomerus,
``Representation Theory of Chern-Simons Observables,''
Duke Math. J.  85 (1996), no. 2, 447-510.
{\tt [arXiv: q-alg/9503016]}


\bibitem{FR}
V.V.Fock, A.A.Rosly,
\newblock{``Poisson structure on moduli of flat connections on Riemann surfaces and $r$-matrix'',}
\newblock{[arXiv: math.QA/9802054]}
  

\bibitem{BNR} E.Buffenoir, K.Noui, Ph.Roche,  {'' Hamitonian Quantization of $SL(2,C)$ Chern-Simons Theory.''}
 Class.Quant.Grav. {\bf 19} (2002) 4953.
{\tt [arXiv:hep-th/0202121]}.

  

\bibitem{BRpreparation}E.Buffenoir, Ph.Roche, ``Chern-Simons Theory with Sources and\\  
Dynamical Quantum Groups II: Unitary representations.'' 
    In preparation.



\bibitem{Harmonic} E. Buffenoir, Ph. Roche,  
``Harmonic analysis on the quantum Lorentz group'',Comm.Math.Phys,{\bf 207}, 499-555, (1999).
{\tt [arXiv: q-alg/9710022]}



\bibitem{ABRR}
 D. Arnaudon, E. Buffenoir, E. Ragoucy, Ph. Roche,
``Universal Solutions of Quantum Dynamical Yang-Baxter Equations'',
Lett.Math.Phys. 44 (1998) 201-214, {[arXiv: \tt q-alg/9712037]}.
%%CITATION = GR-QC 0305079 ; %%

\bibitem{JKOS}
 M. Jimbo, H. Konno, S. Odake, J. Shiraishi,``Quasi-Hopf twistors for elliptic quantum groups'',
{\tt [arXiv: q-alg/9712029 ]}.

\bibitem{Ba}
O.Babelon,`` Universal Exchange Algebra for Bloch Waves and Liouville Theory, '',
{Commun.Math.Phys.139:619-643,1991}.

\bibitem{KM}
P.P.Kulish and A.I.Mudrov,
``Dynamical reflection equations'',
{\tt [arXiv: math.QA/0405556] }



\end{thebibliography}

\section*{Appendix 1: Conventions on $U_q(\mathfrak{G})$}
Let  $\mathfrak{G}$ be a finite dimensional simple complex Lie algebra and denote $\mathfrak{H}$ a Cartan subalgebra of  $\mathfrak{G}.$ 
We denote $\Phi\subset\mathfrak{H}^{*} $ the set of roots of $\mathfrak{G}$. We select a set of simple roots $\Pi=\{\alpha_1,..,\alpha_r\}$ and we denote by $\Phi^+$ the set of positive roots. Let ${\mathfrak G}={\mathfrak N}_{-}\oplus {\mathfrak H}\oplus {\mathfrak N}_{+}$ be the corresponding decomposition of $\mathfrak G$ and let
 ${\mathfrak G}_{\alpha}$ be the root subspace of $\mathfrak G$. 
We denote by $\langle\cdot,\cdot\rangle$ the Killing form on $\mathfrak G.$

If $\alpha\in\mathfrak{H}^{\star}$
we will denote by $t_{\alpha}\in \mathfrak{H}$ the element defined by
$\langle t_{\alpha},h\rangle=\alpha(h),\forall h\in \mathfrak{H},$ and still by 
$\langle \cdot,\cdot\rangle$ the scalar product on $\mathfrak{H}^{\star}$
defined  by duality.
To each root $\alpha$ we will associate the element 
$h_{\alpha}=\frac{2}{\langle \alpha\vert \alpha\rangle}t_{\alpha}.$

A presentation of ${U}_q(\mathfrak{G})$ by generators and
relations is given by:
\begin{eqnarray}
  {[t_{\alpha_i},t_{\alpha_j}]}\  =\  0 &&
  {[e_{\alpha_i},e_{-\alpha_j}]}\  =\  \delta_{ij}\
  \frac{q^{t_{\alpha_i}}-q^{-t_{\alpha_i}}}{q-q^{-1}}
\label{defUq-1}\\
  \begin{array}{l}
  {[t_{\alpha_i},e_{\alpha_j}]} \ =\  a^{\mbox{\tiny sym}}_{ij}
  e_{\alpha_j}\\
  {[t_{\alpha_i},e_{-\alpha_j}]}\ =\ -a^{\mbox{\tiny sym}}_{ij} e_{-\alpha_j} \\
  \end{array}
  &\mbox{ with }& a^{\mbox{\tiny sym}}_{ij}=\langle\alpha_i|\alpha_j\rangle\\
  \begin{array}{l}
    ({\rm ad}_{q'} e_{\alpha_i})^{n_{ij}}(e_{\alpha_j})\ =\ 0\\[.21cm]
    ({\rm ad}_{q'} e_{-\alpha_i})^{n_{ij}}(e_{-\alpha_j})\ =\ 0
  \end{array}
  & \mbox{ if }& i\neq j \mbox{ and }
  n_{ij}=1-2\frac{a^{\mbox{\tiny sym}}_{ij}}{a^{\mbox{\tiny sym}}_{ii}},\ 
  q'=q \mbox{ or } q^{-1}\label{defUq-2}
\end{eqnarray}
where we have introduced:
\begin{equation}
({\rm ad}_{q^{\pm1}} x) (y)= \sum_{(x)} x_{(1)}\, y\, S^{\pm1}(x_{(2)}).
\end{equation}

The Hopf algebra structure is defined by:
\begin{eqnarray}
  \Delta(t_{\alpha_i}) &=& t_{\alpha_i}\otimes 1+1\otimes t_{\alpha_i}\nonumber\\
  \Delta(e_{\alpha_i})&=& e_{\alpha_i}\otimes q^{t_{\alpha_i}}\nonumber
  +1\otimes e_{\alpha_i}\\
  \Delta(e_{-\alpha_i}) &=& e_{-\alpha_i}\otimes 1
  +q^{-t_{\alpha_i}}\otimes e_{-\alpha_i}.
\end{eqnarray}

As usual, we denote by $U_q(\mathfrak{B}_+)$
(resp. $U_q(\mathfrak{B}_-)$) the
algebra generated by $h_{\alpha}, e_{\alpha},\alpha\in\Pi.$
(resp. $h_{\alpha}, e_{-\alpha}, \alpha\in \Pi).$ 

We denote $\pi^+$ (resp. $\pi^-$) the projections of $U_q(\mathfrak{B}_+)$
(resp. $U_q(\mathfrak{B}_-)$) onto $U_q({\mathfrak{ H}})$ . We define  $U^+_q({\mathfrak G})$
(resp. $U^-_q({\mathfrak G})$) the kernel of $\pi^+$ (resp. $\pi^-$).

$\mathfrak{U}_q(\mathfrak{G})$ is quasi-triangular, i.e. there exists an $R-$matrix which obeys the standard quasitriangularity equations
\begin{eqnarray}
(\Delta \otimes id)(R)=R_{13}R_{23}
&&(id \otimes \Delta)(R)=R_{13}R_{12}\label{quasitriangularity}\\
R \Delta &=&\Delta' R \label{quasitriangularity2}.
\end{eqnarray}
As usual we denote $R^{(+)}=R, R^{(-)}=R_{21}^{-1}.$

As a remark, the expression  of $R$  for $U_q(sl(2))$ is simply 
$R=q^{\frac{h\otimes h}{2}}\exp_q((q-q^{-1})e\otimes f)$ with the $q-$exponential being defined as:
\begin{equation}
\exp_q(z)=\sum_{n=0}^{+\infty}\frac{z^n}{(n)_q !}
\end{equation}
where $(n)_q !=(n)_q\cdots (1)_q,$ with $(z)_q=q^{z-1}[z]_q$ and $[z]_q=\frac{q^z-q^{-z}}{q-q^{-1}}.$

 $U_{q}({\mathfrak G})$ is a ribbon Hopf algebra, which means
that it exists an invertible element $v\in U_{q}({\mathfrak G})$ such that:
\begin{eqnarray}
&&\mbox{$v$ is a central element},\nonumber\\
&& v^2=uS(u), \epsilon(v)=1, S(v)=v, \label{v}\nonumber\\
&&\Delta(v)=(R_{21}R_{12})^{-1}(v\otimes v),\label{propertyv}
\end{eqnarray}
where we  have denoted $u=\sum_i S(b_i)a_i$ where $S$ is the antipode and $R=\sum_i a_i \otimes b_i.$ A fundamental property of $u$ is
\begin{eqnarray}
S^2(x)=uxu^{-1},\;\; \forall x \in U_{q}({\mathfrak G}).
\end{eqnarray}

The explicit values of these elements as well as the conventions concerning Clebsch-Gordan coefficients for finite
dimensional representations of $U_q(sl(2))$ are summarized in \cite{Harmonic}.

\section*{Appendix 2: Real forms and Star structures}
 A real vector space $V$ can be equivalently described in terms of its complexification $V^{\mathbb C}=V \otimes_{\mathbb R} {\mathbb C}$ equipped with a star structure, i.e. an antilinear involution, denoted $\star$ and chosen such that $V$ is the real vector space of elements $x\in V^{\mathbb C} $ such that $x^\star=-x.$ 
A complex bilinear form $\langle .,.\rangle$ on  $V^{\mathbb C}$ is the complexification of  a real bilinear form on  $V$ if an only if
 $\overline{\langle x,y\rangle}=\langle x^\star,y^\star\rangle, \forall x,y\in V^{\mathbb C}.$

If $V={\mathfrak g}$ is a Lie algebra we have moreover to require   the star to be an antimorphism of ${\mathbb C}-$Lie algebra.
A real form of a complex Hopf algebra is defined by choosing a star structure which is an antilinear involutive antimorphism of algebra and which property with respect to the coalgebra is usually chosen as being a morphism or antimorphism of coalgebra (see the review in  \cite{Harmonic}.)

 Once a star $\star$ is defined on ${\mathfrak G},$ a star (denoted with the same symbol) can then be straightforwardly defined on $F(G^{\mathbb C})$ by the following requirement added to the definitions (\ref{defM1},\ref{defM2}):
\begin{equation}
({\star \otimes \star})(M)=M^{-1}
\end{equation}
 
We recall here the different real form of $sl(2,{\mathbb C})$ namely $su(2), su(1,1), sl(2,\mathbb{R}).$ Note that although classically $su(1,1)$ is isomorphic as a real Lie algebra to $sl(2,\mathbb R)$, in the quantum case one obtains two different real forms.
These structures are simpler to describe using the following linear involutive automorphism
$\sigma_1,\sigma_2:sl(2,\mathbb{C})\rightarrow sl(2,\mathbb{C})$ defined by
\begin{eqnarray}
&&\sigma_1(h)=h\;\;\;\;\;\;\sigma_1(e)=-e\;\;\;\;\;\;\sigma_1(f)=-f\\
&&\sigma_2(h)=-h\;\;\;\;\;\;\sigma_2(e)=-f\;\;\;\;\;\;\sigma_2(f)=-e.
\end{eqnarray}

We can now define the following star structures on the Lie algebra:
\begin{eqnarray}
&&su(2):e^\star=f\;\;\;f^\star=e\;\;\;h^\star=h,\label{starsu2}\\
&&su(1,1):\overline{\star}=\sigma_1 \circ \star \label{starsu11}\\
&&sl(2,{\mathbb R}):\underline{\star}=\sigma_2 \circ \star.\label{starsl2r}
\end{eqnarray}

$\mathfrak{g}=sl(2,\mathbb{C})_{\mathbb R}$ is the real Lie algebra of $sl(2,\mathbb{C}).$ 
Its complexification  is such that $\mathfrak{g}^{\mathbb{C}}=sl(2,\mathbb{C})\oplus sl(2,\mathbb{C}).$ 
Let us define $e^{(l,r)},f^{(l,r)},h^{(l,r)}$ to be a Cartan basis of $\mathfrak{g}^{\mathbb{C}}$ where
the $l$ (resp. $r$) generate the first (resp.second) component of the direct sum. Any of the following star structure
\begin{eqnarray}
&&(e^{(l)})^\bigstar=f^{(r)}\;\;\;\;(f^{(l)})^\bigstar=e^{(r)}\;\;\;\;(h^{(l)})^\bigstar=h^{(r)}\label{starsl2c1}\\
&&(e^{(l)})^{\overline{\bigstar}}=-f^{(r)}\;\;\;\;(f^{(l)})^{\overline{\bigstar}}=-e^{(r)}\;\;\;\;(h^{(l)})^{\overline{\bigstar}}=h^{(r)}\label{starsl2c2}\\
&&(e^{(l)})^{\underline{\bigstar}}=-e^{(r)}\;\;\;\;(f^{(l)})^{\underline{\bigstar}}=-f^{(r)}\;\;\;\;(h^{(l)})^{\underline{\bigstar}}=-h^{(r)}\label{starsl2c3}
\end{eqnarray}
is a real form on $sl(2,\mathbb{C})\oplus sl(2,\mathbb{C})$ selecting the same real Lie 
algebra $sl(2,\mathbb{C})_{\mathbb R}.$

\section*{Appendix 3: Dynamical Quantum Groups }
The purpose of dynamical quantum group theory (for a review see \cite{ES}) is to give an algebraic framework to the study of solutions of the {\it Dynamical Quantum Yang Baxter Equation } i.e: 
\begin{equation}
R_{12}(\mu)R_{13}(\mu+h_2)R_{23}(\mu)=
R_{23}(\mu+h_1)R_{13}(\mu)R_{12}(\mu+h_3),\label{qdybe}
\end{equation}
where $R:\mathfrak{H}^{*}\rightarrow U_q(\mathfrak{G})^{\otimes 2}$ and $[
R_{12}(\mu),h\otimes 1+1\otimes h]=0, \forall h \in {\mathfrak H}.$

This last equation is a quantization of the classical dynamical Yang Baxter equation, i.e $R$ satisfies $R({\mu})=1+i\hbar r(-\tilde{\chi})+o(\hbar)$ with  ${\mu}\sim -\frac{\tilde{\chi}}{i \hbar}$ and 
$r$ satisfies the Classical Dynamical Yang Baxter quation, i.e:
\begin{eqnarray}
&&[r_{12}(\tilde{\chi}), r_{13}(\tilde{\chi})+r_{23}(\tilde{\chi})]+
[r_{13}(\tilde{\chi}),r_{23}(\tilde{\chi})]=\nonumber\\
&&=
\sum_{i}(h_{\alpha_i}^{(1)}
\frac{\partial r_{23}(\tilde{\chi})}{\partial {\tilde{\chi}_{\alpha_i}}}
-h_{\alpha_i}^{(2)}\frac{\partial r_{13}(\tilde{\chi})}{\partial {\tilde{\chi}_{\alpha_i}}}+h_{\alpha_i}^{(3)}\frac{\partial r_{12}(\tilde{\chi})}{\partial {\tilde{\chi}_{\alpha_i}}}).
\end{eqnarray}

When $r_{12}(\tilde{\chi})$ is the rational solution (\ref{rationalsolutionr}),
it has been shown that the quantization  $R(\mu)$ lies in $U(\mathfrak{G})^{\otimes 2}$ and can  be expressed as:
$$R(\mu)=F_{21}(\mu)^{-1}F_{12}(\mu),$$
where $F(\mu)\in U(\mathfrak{G})^{\otimes 2}$ is the solution of the dynamical 2-cocycle equation
\begin{equation}
(id\otimes \Delta)(F(\mu))F_{23}(\mu)=
( \Delta\otimes id )(F(\mu))F_{12}(\mu+h_{3}),
\end{equation}
with $[
F_{12}(\mu),h\otimes 1+1\otimes h]=0, \forall h \in {\mathfrak H}.$

It has been shown in \cite{ABRR,JKOS} that the corresponding solution of the previous  equation is also the  unique solution of the following linear equation on $F$
\begin{equation}
[F_{12}(\mu),  b(\mu)_2]=-2
(\sum_{\alpha\in\Phi^+}e_\alpha\otimes e_{-\alpha}) F_{12}(\mu)\label{ABRRrat}
\end{equation}
where $b(\mu)=\sum_j (2\mu_j+h_{\alpha_j}) \lambda^j$
 with the condition that
$F(\mu)-1\in U^+(\mathfrak{G})\otimes U^-(\mathfrak{G}).$

When $r_{12}^{\theta}(\tilde{\chi})$ is the trigonometric  solution (\ref{rtrigosolution}),
it has been shown that the quantization  $R^{\theta}(\mu)\in U_q(\mathfrak{G})^{\otimes 2}$ can be expressed as:
\begin{eqnarray}
&&R^{\theta}(\mu)=F^{\theta}_{21}(\mu)^{-1}R^{\theta}F^{\theta}_{12}(\mu)
\end{eqnarray}
where $F^{\theta}(\mu)\in U_{q}(\mathfrak{G})^{\otimes 2}$ is the solution of the dynamical 2-cocycle equation
\begin{equation}
(id\otimes \Delta)(F^{\theta}(\mu))F^{\theta}_{23}(\mu)=
( \Delta\otimes id )(F^{\theta}(\mu))F^{\theta}_{12}(\mu+h_{3}),
\end{equation} $ q=exp(\frac{i\hbar}{4\theta})$ and $R^{\theta}$ is the universal $R$
matrix of $U_q({\mathfrak G}).$
It has been shown in \cite{ABRR,JKOS} that the corresponding solution of the dynamical 2-cocycle equation
 is also the  unique solution of the linear equation on $F^{\theta}$ 
\begin{equation}
F^{\theta}(\mu)B_2(\mu)=R_{12}^{-1}q^{\sum_j h_{\alpha_j}\otimes \lambda^j}B_2(\mu)
F^{\theta}(\mu),
\end{equation}
with $B(\mu)=q^{\sum_j (2\mu_j+h_{\alpha_j}) \lambda^j}=q^{b(\mu)},$ with the condition that
$F^{\theta}(\mu)-1\in U_q^+(\mathfrak{G})\otimes  U_q^-(\mathfrak{G})$ and $[
F_{12}(\mu),h\otimes 1+1\otimes h]=0, \forall h \in {\mathfrak H}.$

This implies the following equation on the dynamical $R$ matrix:
\begin{equation}
R_{12}(\mu)B_2(\mu)R_{21}(\mu)=B_2(\mu+h_1).\label{ABRR2}
\end{equation}

\section*{Appendix 4: Miscellaneous Results}
We compute the Dirac bracket between $V_{\gamma}$ and $V_{\gamma'}$ and obtain the results (\ref{poissonbracketofVr},
\ref{expressionofrtrigodynamique}).

\begin{eqnarray*}
\{V_{\gamma 1},V_{\gamma' 2}\}_{D}&=&\{V_{\gamma 1},V_{\gamma' 2}\}_{d}+2\theta\int \! \! \! \! \! \int _{{\mathbb{g}^{\bot}}\times {\mathbb{g}^{\bot}}}[\mathcal{D}u][\mathcal{D}v]\{V_{\gamma 1},\tilde{\Omega }(u)\}_{d}({\cal K}{}^{\tilde{X}})^{-1}(u,v)\{\tilde{\Omega }(v),V_{\gamma' 2}\}_{d}\\
&=&V_{\gamma 1}V_{\gamma' 2}\left( r_{12}(-\tilde{\chi}_{(k)})-\frac{1}{2\theta}\int \! \! \! \! \! \int _{{\mathbb{g}^{\bot}}\times {\mathbb{g}^{\bot}}}[\mathcal{D}u][\mathcal{D}v]u(\varphi)_1 ({\cal K}{}^{-\frac{1}{4\pi \theta}\tilde{\chi}_{(k)}})^{-1}(u,v)v(\varphi')_2\right)
\end{eqnarray*}
The last identity is obtained using basic properties already established, as well as the following identities
\begin{eqnarray*}
&&\{V_{\gamma 1},V_{\gamma' 2}\}_{d}=V_{\gamma 1}V_{\gamma' 2} r_{12}(-\tilde{\chi}_{(k)})\;\;\;,\;\;\;
\{\tilde{\Omega}(u),M_{(k)}\}_d=0\;\;\;
,\;\;\;\{\tilde{\Omega}(u),A_i\}_d=\frac{1}{2\theta}D_{A_i}u,\\
&& \int _{{\mathbb{g}^{\bot}}\times {\mathbb{g}^{\bot}}}[\mathcal{D}u][\mathcal{D}v] M_{{(k)} 1}^{-1}u(\varphi)_1M_{{(k)} 1} ({\cal K}{}^{X_{(k)}})^{-1}(u,v)M_{{(k)} 2}^{-1}v(\varphi')_2M_{{(k)} 2}=\\
&&
=\int _{{\mathbb{g}^{\bot}}\times {\mathbb{g}^{\bot}}}[\mathcal{D}u][\mathcal{D}v] u(\varphi)_1 ({\cal K}{}^{-\frac{1}{4\pi \theta}\tilde{\chi}_{(k)}})^{-1}(u,v)v(\varphi')_2.
\end{eqnarray*}
This Dirac bracket can also be written as a dynamical quadratic Poisson bracket as
\begin{eqnarray}
\{V_{\gamma 1},V_{\gamma' 2}\}_{D}=V_{\gamma 1}V_{\gamma' 2} r^{\theta}_{12}
(\varphi-\varphi';-\tilde{\chi}_{(k)})
\end{eqnarray}
with
$ r^{\theta}_{12}(\varphi;\tilde{\chi})$ given by:

\begin{equation}
r^{\theta}_{12}(\varphi;-\tilde{\chi})=\frac{1}{4\pi \theta}\left( (\pi-\varphi)\sum_{j}h_{\alpha_j}\otimes \lambda^j+\sum_{\alpha\in \Phi}
e_\alpha\otimes e_{-\alpha}\frac{\pi e^{{\tilde{\chi}}(\alpha)(\pi-\varphi)/4\pi\theta}}{\sinh ({\tilde{\chi}}(\alpha)/4\theta)}\right).
\end{equation}

This is obtained by  computing  explicitely $({\cal K}{}^{Y})^{-1}.$ 
Let $u,v\in  {\mathbb{g}^{\bot}}$ we define 
$u(\varphi)=\sum_{n,\alpha\in \Phi}u_n^\alpha e^{i n\varphi}+
\sum_{n,j=1...r}u_n^{h_{\alpha_j}} e^{i n\varphi},$
with $u_n^\alpha\in {\mathbb C}e_{\alpha}$ and 
$u_n^{h_{\alpha_j}}\in  {\mathbb C}h_{\alpha_j},$ and similarly $v(\varphi)=\sum_{n,\alpha\in \Phi}v_n^\alpha e^{i n\varphi}+
\sum_{n,j=1...r}v_n^{\lambda^j} e^{i n\varphi},$
with $v_n^\alpha\in {\mathbb C}e_{\alpha}$ and 
$v_n^{\lambda^j}\in  {\mathbb C}\lambda^j.$

Let $Y\in {\mathfrak h}$ we have
\begin{eqnarray}
{\cal K}^{Y}(u,v)&=&\langle u,\partial_\varphi v+[Y,v]\rangle\nonumber\\ 
&=&2\pi \left( \sum_{n\not=0,\alpha\in \Phi}\langle u^{-\alpha}_{-n},v^{\alpha}_{n}\rangle(\alpha(Y)+in)+\sum_{n\not=0,j}\langle u^{h_{\alpha_j}}_{-n},v^{\lambda^j}_{n}\rangle\;\; i n\right) .
\end{eqnarray}
As a result we get:
\begin{eqnarray}
({\cal K}^{Y}){}^{-1}(u,v) 
=\frac{1}{2\pi} \left( \sum_{n\not=0,\alpha\in \Phi}\frac{\langle u^{\alpha}_{n},v^{-\alpha}_{-n}\rangle}{\alpha(Y)+in}+\sum_{n\not=0,j}\frac{\langle u^{h_{\alpha_j}}_{n},v^{\lambda^j}_{-n}\rangle}{ i n}\right) .
\end{eqnarray}
Therefore we obtain:
\begin{eqnarray}
&&4\pi\theta r^{\theta}_{12}(\varphi;-\tilde{\chi})=
\sum_{m\not=0,\alpha\in \Phi}
\frac{e_\alpha\otimes e_{-\alpha}}{\frac{1}{4\pi \theta}\tilde{\chi}(\alpha)+i m}e^{im\varphi}+\sum_{j,m\not=0}\frac{h_{\alpha_j}\otimes \lambda^j}{i m}e^{im\varphi}\nonumber\\
&&=(\pi-\varphi)\sum_{j}h_{\alpha_j}\otimes \lambda^j+\sum_{\alpha\in \Phi}
e_\alpha\otimes e_{-\alpha}\frac{\pi e^{{\tilde{\chi}}(\alpha)(\pi-\varphi)/4\pi\theta}}{\sinh ({\tilde{\chi}}(\alpha)/4\theta)},\label{rtrigosolution}
\end{eqnarray}
where in the second equality we have extended the function by $2\pi$-periodicity with the convention that   $\varphi\in[0,2\pi[.$

We now show the result (\ref{DeltaonPquantique},\ref{CommutatorofP}).

We denote $k_{12}=\sum_{j}h_{\alpha_j}\otimes \lambda^j,$ and $a_{12}=\sum_{\alpha\in\Phi^+}e_{\alpha}\otimes e_{-\alpha}.$

\begin{eqnarray*}
\frac{2}{i\hbar} P_1 M_2&=&M_1 b_1(\hat{\mu})M_1^{-1}M_2\\
&=&M_1b_1(\hat{\mu})M_2 R_{12}^{-1}(\hat{\mu})M_1^{-1}\\
&=&M_1M_2(b_1(\hat{\mu})+2k_{12})R_{12}^{-1}(\hat{\mu})M_1^{-1}\\
&=&M_2M_1 R_{12}(\hat{\mu})(b_1(\hat{\mu})+2k_{12})R_{12}^{-1}(\hat{\mu})M_1^{-1}.
\end{eqnarray*}
Using the identity $R_{12}(\hat{\mu})(b_1(\hat{\mu})+2k_{12})R_{12}^{-1}(\hat{\mu})=2t_{12}+b_{1}(\hat{\mu}),$ which is
 a consequence of the linear equation \ref{ABRRrat}, we obtain: 
\begin{equation}
\frac{2}{i\hbar} P_1 M_2=2t_{12}M_2+\frac{2}{i\hbar} M_2 P_1.
\end{equation}

We now prove that $\Delta(P)=P_1+P_2.$ Let ${\cal P}=\frac{1}{2} M b(\hat{\mu})M^{-1},$
\begin{eqnarray*}
2\Delta({\cal P})&=&M_1M_2 F_{12}^{-1}\Delta(b(\hat{\mu}))F_{12}(\hat{\mu})M_2^{-1}M_1^{-1}\\
&=&M_1M_2\Delta(b(\hat{\mu}))M_2^{-1}M_1^{-1}\\
&=&M_1M_2(b_1(\hat{\mu})+b_2(\hat{\mu})+2 k_{12})M_2^{-1}M_1^{-1}\\
&=&M_1b_1(\hat{\mu})M_1^{-1}+M_1M_2b_{2}(\hat{\mu})M_{2}^{-1}M_{1}^{-1}\\
&=&M_1b_1(\hat{\mu})M_1^{-1}+M_2M_1R_{12}(\hat{\mu})b_2(\hat{\mu})
R_{12}^{-1}(\hat{\mu})M_1^{-1}M_{2}^{-1}\\
&=&M_1b_1(\hat{\mu})M_1^{-1}+M_2M_1(b_2(\hat{\mu})+2k_{12}-2t_{12})M_{1}^{-1}M_{2}^{-1}\\
&=&M_1b_1(\hat{\mu})M_1^{-1}+M_2b_2(\hat{\mu})M_2^{-1}-2 M_2M_1 t_{12}M_1^{-1}M_2^{-1}.
\end{eqnarray*}
This closes the proof of  $\Delta(P)=P_1+P_2.$

We now prove the $q-$analog of the two previous relations:
\begin{eqnarray*}
{\cal P}_2 {\breve{R}}_{12}^{-1} U_1&=&(U_2 v_2^{-1} B_2(\hat{\mu}) U_2^{-1}){\breve{R}}_{12}^{-1}U_1=\\
&=&U_2 v_2^{-1} B_2(\hat{\mu}) U_1    R_{21}^{-1}(\hat{\mu})   U_2^{-1}=\\
&=&U_2 U_1 v_2^{-1} B_2(\hat{\mu}+h_1)  R_{21}^{-1}(\hat{\mu})  U_2^{-1}=\\
&=&U_2 U_1 v_2^{-1}R_{12}(\hat{\mu}) B_2(\hat{\mu}) U_2^{-1}=\\
&=&{\breve{R}}_{21}U_1 U_2 v_2^{-1} B_2(\hat{\mu}) U_2^{-1}=\\
&=&{\breve{R}}_{21}U_1 {\cal P}_2
\end{eqnarray*}
and
\begin{eqnarray*}
J_{12}^{-1}\Delta({\cal P})J_{12}&=&J_{12}^{-1}\Delta(U)v_1^{-1}v_2^{-1}R_{21}R_{12}B_2(\hat{\mu})B_1(\hat{\mu}+h_2) \Delta(U^{-1})J_{12}=\\
&=&J_{12}^{-1}
R_{12}^{-1}J_{21} U_1 U_2 F_{12}^{-1}(\hat{\mu})v_1^{-1}v_2^{-1}R_{21}R_{12}B_2(\hat{\mu})B_1(\hat{\mu}+h_2) F_{12}(\hat{\mu}) U_2^{-1} U_1^{-1}J_{21}^{-1}R_{12}
J_{12}=\\
&=&U_2 U_1 R_{12}^{(-)}(\hat{\mu})  F_{12}^{-1}(\hat{\mu})v_1^{-1}v_2^{-1}R_{21}R_{12}B_2(\hat{\mu})B_1(\hat{\mu}+h_2) F_{12}(\hat{\mu}) U_2^{-1} U_1^{-1}{\breve{R}}_{12}=\\
&=&U_2 U_1  F_{21}^{-1}(\hat{\mu}) v_1^{-1}v_2^{-1}R_{12} B_2(\hat{\mu})B_1(\hat{\mu}+h_2) F_{12}(\hat{\mu}) U_2^{-1} U_1^{-1}{\breve{R}}_{12}=\\
&=&U_2 U_1  v_1^{-1}v_2^{-1}R_{12}(\hat{\mu}) B_2(\hat{\mu})B_1(\hat{\mu}+h_2)  U_2^{-1} U_1^{-1}{\breve{R}}_{12}=\\
&=&U_2 U_1  v_1^{-1}v_2^{-1} B_2(\hat{\mu}+h_1)  R_{21}^{-1}(\hat{\mu}) B_1(\hat{\mu}+h_2)  U_2^{-1} U_1^{-1} {\breve{R}}_{12}=\\
&=&U_2 v_2^{-1} B_2(\hat{\mu}) U_1    R_{21}^{-1}(\hat{\mu})   U_2^{-1}  v_1^{-1} B_1(\hat{\mu}) U_1^{-1} {\breve{R}}_{12}=\\
&=&(U_2 v_2^{-1} B_2(\hat{\mu}) U_2^{-1})   {\breve{R}}_{12}^{-1}   (U_1 v_1^{-1} B_1(\hat{\mu}) U_1^{-1}) {\breve{R}}_{12}=\\
&=&{\cal P}_2 {\breve{R}}_{12}^{-1} {\cal P}_1 {\breve{R}}_{12}.
\end{eqnarray*}

\end{document}